\documentclass[%
 reprint,
 amsmath,amssymb,
 aip,cha,
]{revtex4-1}

\usepackage[T1]{fontenc}
\usepackage{xcolor}
\usepackage{graphicx}%
\usepackage{dcolumn}%
\usepackage{bm}%

\newcommand{\stimes}{{\times}}

\begin{document}

\preprint{APS/123-QED}

\title{Inverse Mechano-Electrical Reconstruction of Cardiac Excitation Wave\\ Patterns from Mechanical Deformation using Deep Learning}%

\author{Jan Christoph}
 \email{jan.christoph@ucsf.edu, jan.christoph@med.uni-goettingen.de}
 \affiliation{Department of Cardiology and Pneumology, University Medical Center G\"ottingen, 37075 G\"ottingen, Germany}
\affiliation{German Center for Cardiovascular Research, Partnersite G\"ottingen, 37075 G\"ottingen, Germany}%
 \affiliation{Max Planck Institute for Dynamics and Selforganization, 37077 G\"ottingen, Germany}%

\author{Jan Lebert}
\affiliation{Department of Cardiology and Pneumology, University Medical Center G\"ottingen, 37075 G\"ottingen, Germany}
\affiliation{German Center for Cardiovascular Research, Partnersite G\"ottingen, 37075 G\"ottingen, Germany}%
 \affiliation{Max Planck Institute for Dynamics and Selforganization, 37077 G\"ottingen, Germany}
 \affiliation{Institute for the Dynamics of Complex Systems, University of G\"ottingen, 37077 G\"ottingen, Germany}%

\date{19 November 2020}%

\begin{abstract}

The inverse mechano-electrical problem in cardiac electrophysiology is the attempt to reconstruct electrical excitation or action potential wave patterns from the heart's mechanical deformation that occurs in response to electrical excitation. 
Because heart muscle cells contract upon electrical excitation due to the excitation-contraction coupling mechanism, the resulting deformation of the heart should reflect macroscopic action potential wave phenomena. %
However, whether the relationship between macroscopic electrical and mechanical phenomena is well-defined and furthermore unique enough to be utilized for an inverse imaging technique, in which mechanical activation mapping is used as a surrogate for electrical mapping, has yet to be determined.
Here, we provide a numerical proof-of-principle that deep learning can be used to solve the inverse mechano-electrical problem in phenomenological two- and three-dimensional computer simulations of the contracting heart wall, or in {\it elastic excitable media}, with muscle fiber anisotropy. 
We trained a convolutional autoencoder neural network to learn the complex relationship between electrical excitation, active stress, and tissue deformation during both focal or reentrant chaotic wave activity, and consequently used the network to succesfully estimate or reconstruct electrical excitation wave patterns from mechanical deformation in sheets and bulk-shaped tissues, even in the presence of noise and at low spatial resolutions. %
We demonstrate that even complicated three-dimensional electrical excitation wave phenomena, such as scroll waves and their vortex filaments, can be computed with very high reconstruction accuracies of about $95\%$ from mechanical deformation using autoencoder neural networks, and we provide a comparison with results that were obtained previously with a physics- or knowledge-based approach.

\begin{description}
\item[Keywords]
Complex Systems, Deep Learning, Excitable Media, Inverse Imaging, Ultrasound
\end{description}

\end{abstract}

\maketitle

{\bf
The beating of the heart is triggered by electrical activity, which propagates through the heart tissue and initiates heart muscle contractions. Abnormal electrical activity, which induces irregular heart muscle contractions, is the driver of life-threatening heart rhythm disorders, such as atrial or ventricular tachycardia or fibrillation.
Presently, this abnormal electrical activity cannot be visualized in full, as imaging technology that can penetrate the heart muscle tissue and resolve the inherently three-dimensional electrical phenomena within the heart walls has yet to be developed. A better understanding of the abnormal electrical activity and the ability to visualize it non-invasively and in real-time is necessary for the advancement of therapeutic strategies. 
In this paper, we demonstrate that machine learning can be used to reconstruct the electrical activity from the mechanical deformation, which occurs in response to the electrical activity, in a simplified computer model of a piece of the heart wall. Our study suggests that, in the future, machine learning algorithms could be used in combination with high-speed 3D ultrasound, for instance, to determine the hidden three-dimensional electrical wave phenomena inside the heart walls and to non-invasively image heart rhythm disorders or other electromechanical dysfunctions in patients.
}

\section{Introduction}
\label{sec:introduction}
The heart's function is routinely assessed using either electrocardiography or echocardiography.
Both measurement techniques provide complementary information about the heart.
Whereas electrocardiographic imaging \cite{Oster1997,Ramanathan2004,Cuculich2010,Wang2011}, %
or other electrical techniques, such as catheter-based electro-anatomic mapping \cite{Ben-Haim1996,Gepstein1997,Liu1998,Delacretaz2001}, electrode contact mapping \cite{Nash2001,Nanthakumar2004,Nash2006} or high-density electrical mapping \cite{Konings1994,deGroot2010}, %
provide insights into the heart's electrophysiological state, at relatively high spatial and temporal resolutions on the inner or outer surface of the heart chambers,
echocardiography, in contrast, provides information about the heart's mechanical state \cite{DeBoeck2008}, capturing mechanical contraction and deformation that occurs in response to electrical excitation, throughout the entire heart and the depths of its walls.
Therefore, electrical imaging provides information that mechanical imaging does not provide, and vice versa.
The integrated assessment of both cardiac electrophysiology and mechanics\cite{Kroos2015,Maffessanti2020}, either through simultaneous multi-modality imaging or the processing and interpretation of one modality in the context of the other, could greatly advance diagnostic capabilities and provide a better understanding of cardiac function and pathophysiology. In particular, the analysis of cardiac muscle deformation in the context of electrophysiological activity could help to fill in the missing information that is not accessible with current electrical imaging techniques.

\begin{figure}
\centering
\includegraphics[width=0.45\textwidth]{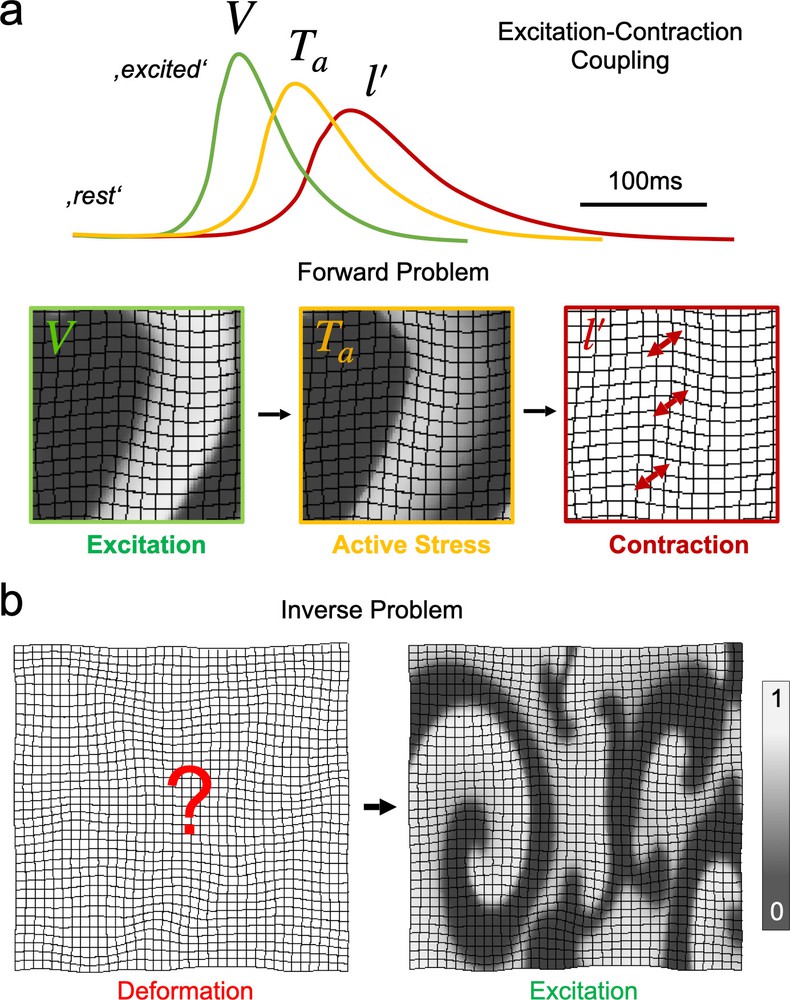}
\caption{Cardiac excitation-contraction coupling (ECC) and the inverse mechano-electrical problem:
\textbf{a)} An action potential triggers intracellular calcium release, which fuels cellular contraction (schematic drawing). The same dynamics with time-course of dynamic variables for excitation $V$, active stress $T_a$ and contractile length change $l'=|1-l/l_0|$ simulated with generic numerical reaction-diffusion-mechanics model (arbitrary units). The 'forward' electro-mechanical problem consists of computing the deformation of a tissue given a spatial excitation and active stress pattern, see section \ref{sec:methods:simulations} for details. Excitation wave pattern travelling from left to right.
\textbf{b)} Inverse mechano-electrical problem in cardiac electrophysiology: reconstruction of excitation wave pattern (right) from mechanical deformation (left). The deformation was caused by the excitation. The deformation can be measured, whereas the excitation can not be measured at all or not in full and needs to be estimated.
The depicted excitation wave pattern (black: resting tissue, white: depolarized or excited tissue) is a generic phenomenological numerical model for cardiac arrhythmias.
}
\label{fig:figure01}
\end{figure}

Because heart muscle cells contract in response to an action potential due to the excitation-contraction coupling mechanism \cite{Bers2002}, see Fig. \ref{fig:figure01}, the resulting deformation on the whole organ level should reflect the underlying electrophysiological dynamics. In anticipation of this, it has been proposed to compute action potential wave patterns from the heart's mechanical deformations using inverse numerical schemes \cite{Otani2010}. 
Furthermore, it was recently demonstrated that cardiac tissue deformation and electrophysiology can be strikingly similar even during heart rhythm disorders \cite{Christoph2018}.
More specifically, it was shown that focal or rotational electrophysiological wave phenomena, such as action potential or calcium waves, visible on the heart surface during ventricular arrhythmias \citep{Davidenko1992,Gray1998,Witkowski1998} induce focal or rotational mechanical wave phenomena within the heart wall, and it was demonstrated that these phenomena can be resolved using high-resolution 4D echocardiography \cite{Christoph2018}. 
The experimental evidence %
supports the notion of electromechanical waves \cite{Wyman1999,Provost2011} that propagate as coupled voltage-, calcium- and contraction waves through the heart muscle \cite{Christoph2018}. 
The high correlation between electrical and mechanical phenomena\cite{Christoph2018,Maffessanti2020} and their wave-like nature\cite{Wyman1999,Provost2011,Christoph2018} has recently motivated the development of an inverse mechano-electrical numerical reconstruction technique\cite{Lebert2019}, that utilizes, in particular, the wave-like nature of strain phenomena propagating through the heart muscle. 
By assimilating observations of these wave-like strain phenomena into a computer model and continuously adapting or synchronizing the model to the observations, such that it develops electrical excitation wave patterns, which in turn cause the model to deform and reproduce the same strain patterns as in the observations, the overall dynamics of the model constitute a reconstruction of the observed dynamics, including the excitation wave dynamics.
Consequently, it was demonstrated {\it in silico} with synthetically generated data that even complicated three-dimensional electrical excitation wave patterns, such as fibrillatory scroll waves and their corresponding vortex filaments, can be reconstructed inside a bulk tissue solely from observing its mechanical deformations.

\begin{figure}
\centering
\includegraphics[width=0.42\textwidth]{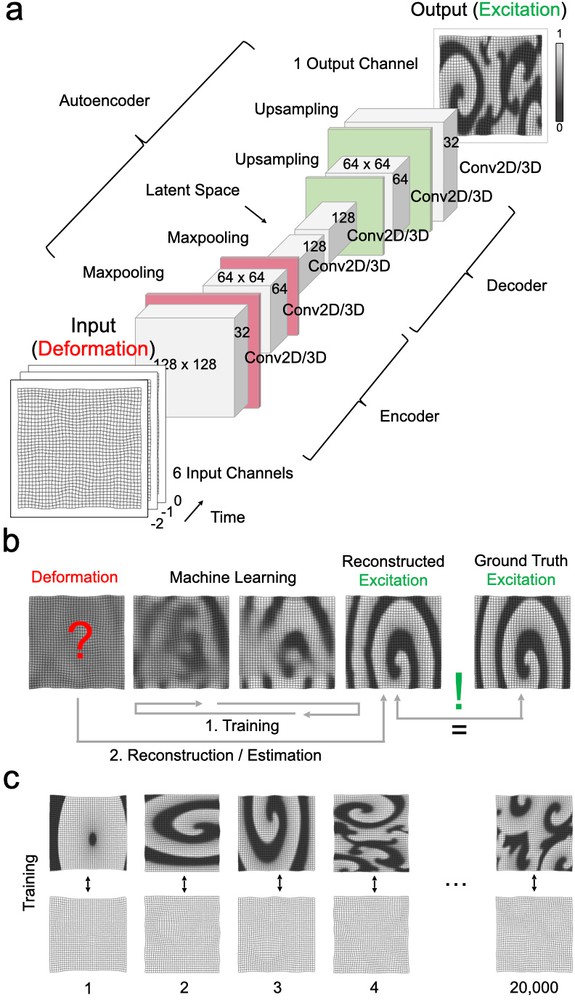}
\caption{Reconstruction of cardiac excitation wave patterns from mechanical deformation using deep learning.
\textbf{a)} Autoencoder neural network for mapping mechanical deformation to electrical excitation patterns. Encoder translates deformation through a series of convolutional and maxpooling layers into latent space, from where it is translated into an excitation pattern through a series of convolutional and upsampling layers (for illustration purposes only 2 stages are shown).
\textbf{b)} During training, the network learns the corresponding distribution of electrical excitation for a given deformation by comparing its obtained reconstruction to a ground truth excitation pattern and adjusting its weights accordingly. 
\textbf{c)} Training with large set of image and video pairs of various electrical excitation and deformation patterns.
}
\label{fig:autoencoder}
\end{figure}

In this study, we present an alternative approach to solving the inverse mechano-electrical problem using deep learning.
The field of machine learning and artificial intelligence has seen tremendous progress over the past decade \cite{LeCun2015}, and techniques such as convolutional neural networks (CNNs) have been widely adopted in the life sciences \cite{Ounkomol2018,Falk2019}. 
Convolutional neural networks \cite{LeCun1989} are a deep learning technique with a particular neural network architecture specialized in learning features in image or time-series data. They have been widely used to classify images \cite{Krizhevsky2012}, segment images \cite{Ronneberger2015}, or recognise features in images, e.g. faces \cite{Lawrence1997}, with very high accuracies, and have been applied in biomedical imaging to segment cardiac magnetic resonance imaging (MRI), computed tomography (CT) and ultrasound recordings \cite{Chen2020}.
Convolutional neural networks are supervised machine learning techniques, as they are commonly trained on labeled data.
Autoencoders are an unsupervised deep learning technique with a particular network architecture, which consists of two convolutional neural network blocks: an encoder and a decoder.
The encoder translates an input into an abstract representation in a higher dimensional feature space, the so-called {\it latent space}, from where the decoder translates it back into a lower dimensional output, and in doing so can generally translate an arbitrary input into an arbitrary output, see Fig. \ref{fig:autoencoder}a).
Convolutional autoencoders employ convolutional neural network layers and are therefore particularly suited to process and translate image data: autoencoders are typically used for image denoising \cite{Vincent2008,Gondara2016}, image segmentation \cite{Ronneberger2015}, image restoration or inpainting \cite{Mao2016}, or for enhancing image resolution \cite{Zeng2017}.
Autoencoders and other machine or deep learning techniques, such as reservoir computing or echo state networks \cite{Jaeger2001,Maas2002,Jaeger2004}, were recently used to replicate and predict chaotic dynamics in complex systems \cite{Pathak2017,Pathak2018,Zimmermann2018,Herzog2018,Herzog2019}.
Particularly, combinations of autoencoders with conditional random fields \cite{Herzog2018}, as well as echo state neural networks \cite{Zimmermann2018}, were recently applied to predict the evolution of electrical spiral wave chaos in excitable media and to "cross-predict" one dynamic variable from observing another.
Moreover, it was recently shown that convolutional neural networks can be used to detect electrical spiral waves in excitable media \cite{Mahesh2020}, and, lastly, deep learning was used to estimate the deformation between two cardiac MRI images \cite{Qiu2020}, and the ejection fraction in ultrasound movies of the beating heart \cite{Ouyang2020}.
The recent progress and applications of deep learning to excitable media and cardiac deformation quantification suggests that deep learning can also be applied to solving the inverse mechano-electrical problem.

Here, we apply an autoencoder-based deep learning approach to reconstruct excitation wave dynamics in elastic excitable media by observing and processing mechanical deformation that was caused by the excitation. 
We developed a neural network with a 2D and 3D convolutional autoencoder architecture,
which is capable of learning mechanical spatio-temporal patterns and translating them into corresponding electrical spatio-temporal patterns.
We trained the neural network on a large set of image- and video-image pairs, showing on one side mechanical deformation and on the other side electrical excitation patterns, to have the network learn the complex relationship between excitation, active stress and deformation in computer simulations of an elastic excitable medium with muscle fiber anisotropy, see Fig. \ref{fig:autoencoder}.
We consequently show that it is possible to predict or reconstruct electrical excitation wave patterns, even complicated two- and three-dimensional spiral or scroll wave chaos, from the deformations that they have caused, and therefore provide a numerical proof-of-principle that the inverse mechano-electrical problem can be solved using machine learning.

\section{Materials and Methods}
\label{sec:MaterialsAndMethods}

We generated two- and three-dimensional synthetic data of excitation waves in an accordingly deforming elastic excitable medium with muscle fiber anisotropy, and used the data to train a convolutional autoencoder neural network, see Fig. \ref{fig:autoencoder}c), to learn the complex relationship between excitation and direct kinematic quantities such as tissue displacements as the medium deforms due to the excitation via the excitation-contraction coupling mechanism, see section \ref{sec:methods:simulations}. The trained network was then used to estimate time-varying spatial distributions of electrical excitation that caused a particular time-varying mechanical deformation, see Fig. \ref{fig:autoencoder}.
The neural network uses static or dynamic images or videos of mechanical deformation as input, respectively, and finds the corresponding distributions of excitation that had caused the deformation and returns them as output. We then compared the estimated excitation to the ground truth excitation. Next to excitation, also active stress can be estimated independently from mechanical deformation.

\begin{figure}
\centering
\includegraphics[width=0.42\textwidth]{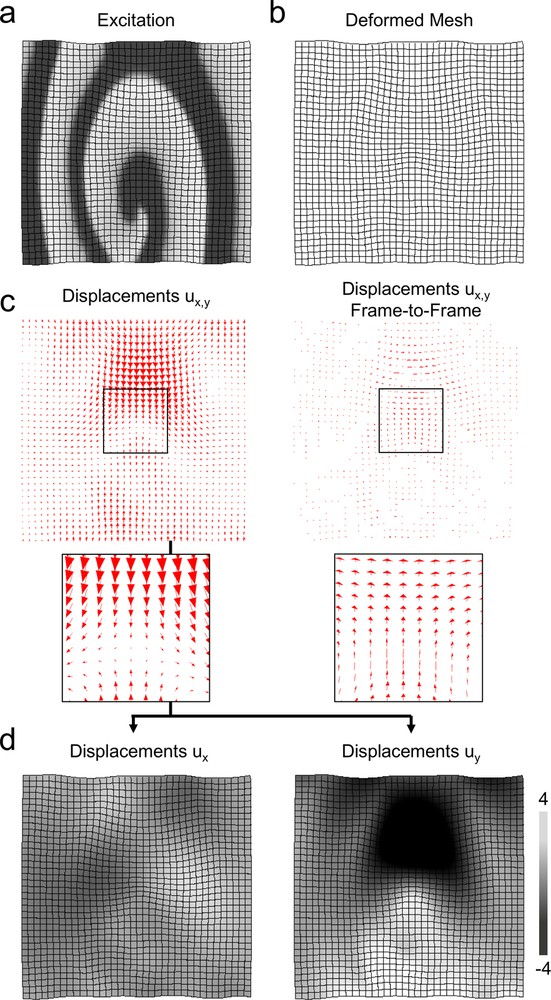}
\caption{Mechanical input data analyzed by neural network:
\textbf{a)} Original excitation pattern inducing deformations.
\textbf{b)} Deformation displayed as deformed mesh.
\textbf{c)} Deformation displayed as displacement vector field. Left: displacement vector field $\vec{u}_r$ showing motion with respect to undeformed reference configuration. Right: Instantaneous, frame-to-frame displacement field $\vec{u}_i$ showing motion of each material coordinate with respect to previous time step, see also Fig. \ref{fig:temporal}c).
\textbf{d)} Magnitudes of individual displacement components $u_x$ and $u_y$ drawn on deformed mesh (units [-4,4] pixels). The $u_x$- and $u_y$-components of either $\vec{u}_r$ or $\vec{u}_i$ are fed as individual input channels into the network, see also Fig. \ref{fig:temporal}.
}
\label{fig:kinematics}
\end{figure}

\begin{figure*}
  \centering
  \includegraphics[width=0.95\textwidth]{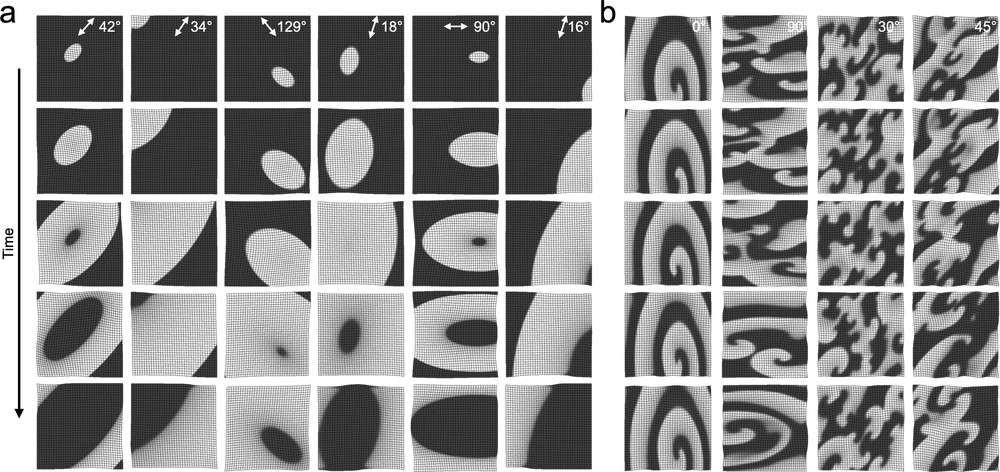}
  \caption{
    Two-dimensional training data for deep-learning-based reconstruction of electrical excitation wave patterns from mechanical deformation.
    \textbf{a)} Focal training data with focal and quasi-planar electrical excitation waves originating from randomly chosen stimulation sites with random electrical parameters or wavelengths and fiber orientations (from $0-360^{\circ}$, white arrows). Dataset includes $200$ simulations of $200$ focal waves and about $50,000$ video frames after data-augmentation.
    \textbf{b)} Spiral and chaotic training data consisting of different episodes showing single spiral waves and persistent and decaying spiral wave chaos. Dataset includes $10$ different simulations and about $50,000$ video frames after data-augmentation.
    Overall the training data includes a variety of dynamical regimes.
    The different trained models and the data used for training are summarized in Table \ref{table:networkarchitecture}. Just focal data as in a), or just chaotic spiral training data as in b), or a mixture of the two datasets was used for training. 
    The training data and the data on which the reconstruction was performed (validation data) are two separate datasets.
  }
  \label{fig:trainingdata}
\end{figure*}

\subsection{Neural Network Architecture}\label{sec:methods:networkarchitecture}
We developed a convolutional autoencoder neural network comprised of convolutional layers, activation layers and maxpooling\cite{Ranzato2007} layers, an encoding stage that uses a two- or three-dimensional vector field of tissue displacements $\vec{u} \in \mathbb{R}^2,\mathbb{R}^3$ as input, and a decoding stage that outputs a two- or three-dimensional scalar-valued spatial distribution of estimates for the excitation $\widetilde{V} \in \mathbb{R}$, respectively, which are approximations of the ground truth excitation $V \in \mathbb{R}$ that had originally caused the deformation. 
Next to the electrical excitation, the neural network can also be trained to estimate the distribution of active stress $T_a \in \mathbb{R}$, see also section \ref{sec:methods:simulations}.
More precisely, the input vector field can optionally either be (i) a static two-dimensional vector field $\vec{u}_{x,y}(x,y)$ describing a deformation in two-dimensional space, (ii) a static three-dimensional vector field $\vec{u}_{x,y,z}(x,y,z)$ describing a deformation in three-dimensional space, or (iii) a time-varying two-dimensional vector field $\vec{u}_{x,y}(x,y,t)$ describing motion and deformation in two-dimensional space. The latter vector fields are given as short temporal sequences of two- or three-dimensional vector fields, respectively, e.g. $\{ \vec{u}_{t_{-2}}, \vec{u}_{t_{-1}},\vec{u}_{t_0} \}$ with the temporal offset $\tau = |t_0-t_{-1}|$ between vector fields being about of $5\%$ of the dominant period of the activity, e.g. the spiral period. 
The vector fields describe displacements of the tissue $\vec{u}_r$ with respect to either the stress-free, undeformed mechanical reference configuration $\chi_0$, or, alternatively, instantaneous shifts $\vec{u}_i$ from frame to frame with respect to either the stress-free, undeformed configuration $\chi_0$ or an arbitrary configuration $\chi_t$, see Fig. \ref{fig:kinematics}c) and Fig. \ref{fig:temporal}c).
The basic network architecture is a convolutional autoencoder with 3 stages in the encoding and decoding parts, respectively, see Fig. \ref{fig:figure01}b). Each stage corresponds to a convolutional layer, an activation layer, and a maxpooling or upsampling layer, respectively.
Table \ref{table:networkarchitecture} summarizes the different autoencoder models used in this work, which all share a similar autoencoder network architecture.
In the following, we distinguish the different autoencoder models by whether they process two-dimensional or three-dimensional data, by whether they process static or time-varying data, as well as by the spatial input resolution and filter numbers of the convolutional layers. In addition, we distinguish the models by the type of data that they were exposed to during training, see section \ref{sec:methodstraining}, Table \ref{table:networkarchitecture} and Fig. \ref{fig:trainingdata}. The notation (s) or (f) at the end of the model description indicates that the model was just trained on spiral chaos or focal wave data. Otherwise it was trained on both data types.

The models for (time-varying) two-dimensional data use padded two-dimensional convolutional layers (2D-CNN) with filter size $3\stimes3$ and rectified linear unit\cite{Nair2010} (ReLU) as activation function. All convolutional layers of models 3Ds-xx are padded three-dimensional convolutions (3D-CNN) with filter size $3\stimes3\stimes3$ followed directly by a batch normalization\cite{Ioffe2015} layer (to accelerate the training and improve the accuracy of the network) and a ReLU activation layer.
Some two-dimensional models (e.g. 2Dt-A3/s/f and/or 2Dt-B3s) read a series of 3 subsequent two-dimensional video images as input, the individual images showing mechanical deformations induced by focal (f) and/or spiral wave chaos (s) at time $t_0$ and two previous timesteps $t_{-1}$ and $t_{-2}$, see also section \ref{sec:results:temporal}. Therefore, note that, if we refer to 'video images / frames' or 'samples', each of these samples may refer to a single or a short series of $2-3$ frames in the case that the network analyzes spatio-temporal input.  
Model 2Ds-A1 reads only a single video image (2 input channels) and therefore processes only static deformation data.
The number of input channels of the first network layer is $n_c = n_d \cdot n_T$ for all models, where $n_d$ is the number of dimensions of each input displacement vector and $n_T$ is the number of time steps that are used for the reconstruction, e.g. $2 \cdot 3 = 6$ input channels corresponds to $u_x$- and $u_y$-components of a two-dimensional vector field (see Fig. \ref{fig:kinematics}d)) for 3 time steps. The spatial input size of the models is given in Table \ref{table:networkarchitecture} and depends on whether spatial subsampling of the dataset is used, see section \ref{sec:results:resolution} and Fig. \ref{fig:resolution}d).
The last network layer for all models is one convolutional filter using a sigmoid activation function, the spatial output size is $128\stimes128$ for 2D-CNNs and $104\stimes104\stimes24$ for 3D-CNNs for all models (see section \ref{sec:methodstraining}). 

For instance, model 2Dt-B3s processes time-varying two-dimensional (2D+t) video data and retains $334,241$ trainable parameters, see Table \ref{table:networkarchitecture}. The network consists of 1 input layer of size $128 \stimes 128 \stimes 6$ with 6 channels for 3 video frames, each frame containing 2 components for $u_x$- and $u_y$-displacements, respectively, and 7 convolutional layers in total. The encoding part consists of a total of 3 convolutional layers of sizes $128 \stimes 128 \stimes 32$, $64 \stimes 64 \stimes 64$ and $32 \stimes 32 \stimes 128$, each followed by a maxpooling layer of filter size $2\stimes2$ reducing the spatial size by a factor of 2. The decoding part of the autoencoder network architecture contains 3 convolutional layers of sizes $16 \stimes 16 \stimes 128$, $32 \stimes 32 \stimes 64$ and $64 \stimes 64 \stimes 32$, each followed by a upsampling layer of filter size $2\stimes2$ increasing the spatial size by a factor of 2, and a final convolutional layer of size $128 \stimes 128 \stimes 1$ with a sigmoid activation function. All networks were implemented in Keras\cite{chollet2015keras} with the Tensorflow\cite{tensorflow2015-whitepaper} backend.

\begin{table*}
\begin{center}
{\small 
\begin{tabular}{|p{1.4cm}|p{1.5cm}|p{3.9cm}|p{0.8cm}|p{2.0cm}|p{2.0cm}|p{1.7cm}|p{3.2cm}| }
 \hline
 Model  & Model Param.   & Convolutional Layers\newline Number of Filters       & Input Ch.  & Input Size                   & Accuracy              & Training Duration & Training Data Type\\ \hline
 2Dt-A3  & 1,332,033         & 64$\downarrow$128$\downarrow$256$\downarrow$--256$\uparrow$128$\uparrow$64$\uparrow$     & 6             & $128 \stimes 128$             &  $96.2\% \pm 3.1\%$   & 15-20 min. & Focal, Spiral, Chaos   \\ 
 2Dt-A3f  & 1,332,033         & 64$\downarrow$128$\downarrow$256$\downarrow$--256$\uparrow$128$\uparrow$64$\uparrow$     & 6             & $128 \stimes 128$             & $98.6\% \pm 1.4\%$    & 15-20 min. & Focal   \\
 2Dt-A3s  & 1,332,033         & 64$\downarrow$128$\downarrow$256$\downarrow$--256$\uparrow$128$\uparrow$64$\uparrow$     & 6             & $128 \stimes 128$             & $95.1\% \pm 2.6\%$    & 15-20 min. & Spiral, Chaos   \\
 2Dt-A2s  & 1,330,881         & 64$\downarrow$128$\downarrow$256$\downarrow$--256$\uparrow$128$\uparrow$64$\uparrow$     & 4             & $128 \stimes 128$             & $94.8\% \pm 2.6\%$    & 15-20 min. & Spiral, Chaos   \\
 2Ds-A1s  & 1,329,729         & 64$\downarrow$128$\downarrow$256$\downarrow$--256$\uparrow$128$\uparrow$64$\uparrow$     & 2             & $128 \stimes 128$             & $93.4\% \pm 3.2\%$    & 15-20 min. & Spiral, Chaos          \\
 2Ds-A1  & 1,329,729         & 64$\downarrow$128$\downarrow$256$\downarrow$--256$\uparrow$128$\uparrow$64$\uparrow$     & 2             & $128 \stimes 128$             & $95.9\% \pm 3.3\%$    & 15-20 min. & Focal, Spiral, Chaos          \\
 2Dt'-A2s  & 1,330,881         & 64$\downarrow$128$\downarrow$256$\downarrow$--256$\uparrow$128$\uparrow$64$\uparrow$     & 4             & $128 \stimes 128$         & $94.8\% \pm 2.6\%$    & 15-20 min. & Spiral, Chaos (inst.)   \\
 2Dt'-A1s  & 1,329,729         & 64$\downarrow$128$\downarrow$256$\downarrow$--256$\uparrow$128$\uparrow$64$\uparrow$     & 2             & $128 \stimes 128$         & $93.5\% \pm 3.2\%$    & 15-20 min. & Spiral, Chaos (inst.)   \\
 2Dt$^*$-A2s  & 1,330,881         & 64$\downarrow$128$\downarrow$256$\downarrow$--256$\uparrow$128$\uparrow$64$\uparrow$     & 4             & $128 \stimes 128$         & $94.8\% \pm 2.6\%$    & 15-20 min. & Spiral, Chaos (ref. $\chi_t$)   \\
 2Ds$^*$-A1s  & 1,329,729         & 64$\downarrow$128$\downarrow$256$\downarrow$--256$\uparrow$128$\uparrow$64$\uparrow$     & 2             & $128 \stimes 128$         & $93.5\% \pm 3.0\%$    & 15-20 min. & Spiral, Chaos (ref. $\chi_t$)   \\
 2Dt-B3s  & 334,241           & 32$\downarrow$64$\downarrow$128$\downarrow$--128$\uparrow$64$\uparrow$32$\uparrow$       & 6             & $128 \stimes 128$             & $94.5\% \pm 2.7\%$    & 5-10 min. & Spiral, Chaos  \\
 2Ds-B1s  & 333,089           & 32$\downarrow$64$\downarrow$128$\downarrow$--128$\uparrow$64$\uparrow$32$\uparrow$       & 2             & $128 \stimes 128$             & $92.4\% \pm 3.5\%$    & 5-10 min. & Spiral, Chaos  \\
 2Dt-C3s  & 229,329           & 64$\downarrow$128$\downarrow$--128$\uparrow$64$\uparrow$       & 6             & $128 \stimes 128$             & $92.7\% \pm 3.4\%$    & 5-10 min. & Spiral, Chaos  \\
 2Dt-D3s  & 741,953           & 128$\downarrow$256$\downarrow$--256$\uparrow$128$\uparrow$64$\uparrow$       & 6             & $64 \stimes 64$             & $94.1\% \pm 3.1\%$    & 5-10 min. & Spiral, Chaos  \\
 2Ds-E1s  & 518,337           & 64$\downarrow$--256$\uparrow$128$\uparrow$64$\uparrow$       & 2             & $32 \stimes 32$             & $89.9\% \pm 4.3\%$    & 5-10 min. & Spiral, Chaos  \\
 2Dt-F3s  & 520,641           & 64--256$\uparrow$128$\uparrow$64$\uparrow$       & 6             & $16 \stimes 16$             & $90.8\% \pm 4.7\%$    & 5-10 min. & Spiral, Chaos  \\
 2Ds-F1s  & 518,337           & 64--256$\uparrow$128$\uparrow$64$\uparrow$       & 2             & $16 \stimes 16$             & $87.3\% \pm 5.0\%$    & 5-10 min. & Spiral, Chaos  \\
 3Ds-A1 & 1,000,131           & 32$\downarrow$64$\downarrow$128$\downarrow$--128$\uparrow$\,64$\uparrow$32$\uparrow$      & 3            & $104 \stimes 104 \stimes 24$  & $95.7\% \pm 0.5\%$    & $2.5$ hours & Chaos  \\
 3Ds-B1 & 1,000,131           & 32-64$\downarrow$128$\downarrow$--128$\uparrow$64$\uparrow$32$\uparrow$            & 3            & $52 \stimes 52 \stimes 12$     & $95.7\% \pm 0.6\%$    & 1-2 hours & Chaos  \\
 3Ds-C1 & 947,331             & 64$\downarrow$128$\downarrow$--128$\uparrow$64$\uparrow$32$\uparrow$         & 3            & $52 \stimes 52 \stimes 12$    & $94.8\% \pm 0.6\%$    & 1-2 hours & Chaos  \\
 3Ds-D1 & 1,000,131           & 32-64-128$\downarrow$--128$\uparrow$64$\uparrow$32$\uparrow$            & 3            & $26 \stimes 26 \stimes 6$     & $94.8\% \pm 0.7\%$    & 1-2 hours & Chaos  \\
 3Ds-E1 & 731,139             & 128$\downarrow$--128$\uparrow$64$\uparrow$32$\uparrow$            & 3            & $26 \stimes 26 \stimes 6$     & $92.6\% \pm 0.9\%$    & 1-2 hours & Chaos  \\
 3Ds-F1 & 1,000,131           & 32-64-128--128$\uparrow$64$\uparrow$32$\uparrow$            & 3            & $13 \stimes 13 \stimes 3$     & $89.4\% \pm 1.4\%$    & 1-2 hours & Chaos  \\
 3Ds-G1 & 288,387           & 128$\uparrow$64$\uparrow$32$\uparrow$            & 3            & $13 \stimes 13 \stimes 3$     & $85.1\% \pm 1.4\%$    & 1-2 hours & Chaos  \\
 \hline
\end{tabular}
}
\caption{Comparison of neural network models with different convolutional autoencoder architectures used for inverse mechano-electrical reconstruction of excitation wave dynamics from mechanical deformation in numerical models of cardiac tissue. The different network models were trained on a Nvidia GeForce RTX 2080 Ti GPU. Maxpooling or upsampling layers are indicated by $\downarrow$ or $\uparrow$, respectively. Note that some of the 2D network models were trained only on focal data (e.g. 2Dt-A3f), some only on spiral chaos data (e.g. 2Dt-A3s, 2Dt'-A1s) and some on a 50\%-50\% mixture of focal and spiral chaos data (e.g. 2Dt-A3, 2Ds-A1), see also Fig. \ref{fig:trainingdata}. Further, some models process static (s) data (e.g. 2Ds-A1 or 3Ds-A1), whereas others process spatio-temporal (t) data (e.g. 2Dt-A3 or 2Dt-F3s). See section \ref{sec:results:temporal} for a description of $'$, $*$ and (inst. /ref. $\chi_t$).} 
\label{table:networkarchitecture}
\end{center}
\end{table*}

\subsection{Training Data Generation using Computer Simulations of Elastic Excitable Media}\label{sec:methods:simulations}

Two- and three-dimensional deformation and excitation wave data was generated using computer simulations. The source code for the computer simulations is available in \cite{Lebert2019}.
In short, nonlinear waves of electrical excitation, refractoriness and active stress were modeled in two- or three- dimensional simulation domains representing the electrical part of an elastic excitable medium using partial differential equations and an Euler finite differences numerical integration scheme, as described previously \cite{AlievPanfilov1996,Nash2004,Panfilov2007}:
\begin{eqnarray}
  \label{eq:modelu}
  \frac{\partial V}{\partial t} & = & \nabla^2 V - k V (V-a) (V-1) - V r - I_s \\
  \label{eq:modelv}
  \frac{\partial r}{\partial t} & = & \epsilon(V,r)(k V(a+1-V)-r) \\
  \label{eq:modelTa}
  \frac{\partial T_a}{\partial t} & = & \epsilon (V)(k_T V - T_a) 
\end{eqnarray}
Here, $V$, $r$ and $T_a$ are dimensionless, normalized dynamic variables for excitation (voltage), refractoriness and active stress, respectively, and $k$ and $a$ are electrical parameters, which influence properties of the excitation waves, such as their wavelengths.
Together with the diffusive term in equation (\ref{eq:modelu}),
\begin{eqnarray}
\label{eq:diffusionconstant}
  \nabla^2 V & = & \nabla \cdot (D \nabla V) 
\end{eqnarray}  
including the diffusion constant $D$, the model produces nonlinear waves of electrical excitation (anisotropic in two-dimensional model) followed by waves of active stress and contraction, respectively. 
The contraction strength or magnitude of active stress is regulated by the parameter $k_T$.

The electrical model facilitates stretch-activated mechano-electrical feedback via a stretch-induced ionic current $I_s$, which modulates the excitatory dynamics upon stretch: 
\begin{eqnarray}
  \label{eq:Gs}
  I_s & = & G_s (\sqrt{A}-1)(V-E_s) 
\end{eqnarray}
The current strength depends on the area $A$ of one cell of the deformable medium, the equilibrium potential (here $E_s=1$), and the parameter $G_s$, which regulates the maximal conductance of the stretch-activated ion channels.
With excitation-contraction coupling, and stretch-activated mechano-electrical feedback, the model is coupled in forward and backward direction.

Soft-tissue mechanics were modeled using a two- or three-dimensional mass-spring damper particle system with tunable fiber anisotropy and a Verlet finite differences numerical integration scheme, similarly as described previously \cite{Bourguignon2000,Mohr2003,Weise2011,Lebert2019}. 
In short, the elastic simulation domain consists of a regular grid of quadratic or hexaedral cells (pixels/voxels), respectively, each cell being defined by vertices and lines or faces, respectively. The edges are connected by passive elastic springs. 
At the center of each cell is a set of two- or three orthogonal springs that can be arbitrarily oriented. One of the springs is an active spring, which contracts upon excitation or active stress and is aligned along a defined fiber orientation.
In the two-dimensional model, muscle fibers are aligned uniformly or linearly transverse and can be aligned in any arbitrary direction ($0-360^{\circ}$).
In the three-dimensional model, muscle fibers are organized in an orthotropic stack of sheets, which are stacked in $e_z$-direction with the fiber orientation rotating by a total angle of $90^{\circ}$ throughout the stack.
The muscle fiber organization leads to highly anisotropic contractions and deformations of the sheet or bulk tissues.
The elastic part of the model exhibits large deformations.
The size of the two- and three-dimensional simulation domains was $200 \stimes 200$ cells/pixels and $100 \stimes 100 \stimes 24$ cells/voxels, respectively, where one electrical cell corresponds to one mechanical cell.

Fig. \ref{fig:figure01}a) shows an example of an excitation wave propagating through an accordingly deforming tissue (from left to right). 
Depolarized or excited tissue is shown in white and repolarized or resting tissue is shown in black (normalized units on the interval [0,1]).
The excitation wave is followed by an active stress wave, which exerts a contractile force along fibers (red arrows). Note that the contraction sets in at the tail of the excitation wave, since there is a short electromechanical delay between excitation and active stress.
The numerical simulation was implemented in C++ and runs on a CPU and uses multi-threading for parallel computation on multiple cores.

\subsection{Training Data and Training Procedure}\label{sec:methodstraining}

Using the computer simulation described in section \ref{sec:methods:simulations}, we generated two- and three-dimensional training datasets consisting of corresponding mechanical and electrical data.
The two-dimensional training data includes two datasets, one with focal wave data and the other with single spiral wave and spiral wave chaos data, see Fig. \ref{fig:trainingdata}a) and b).
The focal dataset includes data from $200$ different simulations of $200$ focal or target waves originating from randomly chosen stimulation sites, as shown in Fig. \ref{fig:trainingdata}a). The simulations were initiated with randomized electrical parameters $a \in [0.05,0.1]$ (in steps of 0.01), $k \in [7,8,9]$, $G_s \in [0,1]$ (in steps of 0.1), and randomized values for the diffusion constant $D \in [0.1,1]$ (in steps of 0.1) and fiber angle $\alpha \in [0^{\circ},90^{\circ}]$ (in steps of $1^{\circ}$).
Each focal wave sequence is comprised of $80$ video frames, which were each saved every $50$ simulation time steps.
Using data-augmentation, the size of the focal training data was increased to approximately $63,000$ frames, and then $20,000$ frames were randomly chosen for training, see also Fig. \ref{fig:training}c).
Data-augmentation obviated computing additional training data in computer simulations.
During data-augmentation, the two-dimensional data was firstly rotated by $90^{\circ}$, and then flipped both horizontally and vertically, such that all fiber alignments from $0-360^{\circ}$ were equally likely to be present in the data.
The spiral wave and spiral wave chaos dataset includes data from $10$ different simulations with either single stationary or meandering spiral waves or persistent or decaying spiral wave chaos, as shown in Fig. \ref{fig:trainingdata}b).
Each of the $10$ datasets was initiated with slightly different electrical and/or mechanical parameters with a fiber alignment of $0^{\circ}$, $15^{\circ}$, $30^{\circ}$ or $45^{\circ}$ (and accordingly $180^{\circ}$, $195^{\circ}$, $210^{\circ}$ and $225^{\circ}$, respectively), and consists of $300-800$ video frames, which were each saved every $50$ simulation time steps.
The $10$ datasets contained in total about $7,000$ video frames.
Using data-augmentation, the size of the spiral chaos training data was increased to approximately $57,000$ frames, and then $20,000$ frames were randomly chosen for training, see also Fig. \ref{fig:training}c).
During data-augmentation, the two-dimensional data was firstly rotated by $90^{\circ}$, such that in addition to the $0^\circ$, $15^\circ$, $30^\circ$ and $45^\circ$ fiber alignments, see section \ref{sec:methods:simulations}, the data also included $90^{\circ}$, $105^{\circ}$, $120^{\circ}$ and $135^{\circ}$ fiber alignments. Then, the data was flipped both horizontally and vertically, such that all fiber alignments from $0-360^{\circ}$ in steps of $15^{\circ}$, including $60^{\circ}$, $75^{\circ}$, $150^{\circ}$ $165^{\circ}$ and their corresponding inverse vectors were present in the training data.
For training on both focal and spiral chaos data, $20,000$ frames were randomly chosen from both datasets with $50\%$ of the frames being focal and $50\%$ being spiral chaos data.
Each video frame of both the focal and spiral chaos datasets was resized from $200 \stimes 200$ cells to $128 \stimes 128$ pixels before being provided as input to the network.

The three-dimensional training datasets contain initially $8,400$ volume frames selected randomly from a set of $11$ different chaotic simulations containing $2,000$ volume frames each.
Using data-augmentation, we increased the size of each three-dimensional training dataset to approximately $67,000$ frames by randomly flipping frames along the $e_x$-, $e_y$- and/or $e_z$-axis.
Validation was performed after training on a different dataset containing $16,500$ volume frames data from $11$ different simulations with $1,500$ volume frames each.
Each training or validation simulation uses a different value for the mechano-electrical feedback strength $G_s \in [0, 2]$ (in steps of $0.2$), see Fig. \ref{fig:results3D1}e), the electrical parameters $a=0.05$, $k=8$, and the diffusion constant $D=0.05$ are the same for all simulations. The fiber angle $\alpha$ rotates linearly between $0^{\circ}$ (bottom) and $90^{\circ}$ (top) of the thin three-dimensional bulk.
The simulation data was padded with zeros from the simulation domain of $100 \stimes 100 \stimes 24$ cells to $104 \stimes 104 \stimes 24$ voxels to fit the network architecture, the neural network predictions are truncated back to $100 \stimes 100 \stimes 24$ voxels for analysis.

The two- and three-dimensional training data was specifically not used during the reconstruction. For validation purposes (e.g. determining the reconstruction error) the two-dimensional reconstructions were performed on a validation dataset including $20,000$ video frames. All validation datasets were separate datasets in order for the network to always exclusively estimate excitation patterns from data that it had not already seen during training. A fraction of 20\% of the frames of the training datasets were used for testing during training.
To simulate noisy mechanical measurement data, we added noise to a fraction of the mechanical training data, see Fig. \ref{fig:noise}.
The noise, normally distributed with varying average amplitudes, was added to the individual displacement vector components independently in each frame and independently over time.

The 2D-CNNs were trained with a batch size of $128$ using the Adadelta\cite{ADADELTA} optimizer with a learning rate of $0.001$ and binary cross entropy as loss function, whereas the 3D-CNNs were trained with a batch size of $16$ using the Adam\cite{Adam} optimizer with a learning rate of $0.001$ and mean squared error as loss function.
All models were typically trained for $50$ epochs, if not stated otherwise, see also Fig. \ref{fig:training}.
With the models 2Dt-xx, the training procedure takes roughly $20$ seconds per epoch or $15-20$ minutes in total with $\sim 20,000$ frames or samples, respectively.
With model 3Ds-A1, the training procedure takes roughly $160$ seconds per epoch or $2.5$ hours in total for $50$ epochs with $8,400$ samples which are augmented during training.
Training and reconstructions were performed on a Nvidia GeForce RTX 2080 Ti graphics processing unit (GPU).

\subsection{Reconstruction Accuracy and Error}
The overall reconstruction error of a neural network model (2D/3D) was determined by calculating the absolute differences between all estimated excitation values $\widetilde{V}$ and original ground truth excitation values $V$:
\begin{eqnarray}
  \label{eq:error}
  \langle \Delta V(x,y,t) \rangle & = & \frac{1}{N} \sum |\widetilde{V}(x,y,t)-V(x,y,t)|
\end{eqnarray}
and calculating the average over all pixels or voxels and all frames (mean absolute error: MAE), respectively. 
The reconstruction accuracy is implicitly given as $1-\langle \Delta V \rangle$.
Throughout this study, the uncertainty of the reconstruction error or accuracy is stated either as i) the standard deviation $\sigma_{\langle V \rangle}$ of the per-frame-average of the reconstruction error across the frames, or ii) the standard deviation $\sigma_V$ of all absolute difference values over all pixels or voxels in all frames (stated in brackets), cf. Fig. \ref{fig:training}b,c) black and grey error bars, respectively.
For two-dimensional data, the reconstruction error was calculated using $N=20,000$ estimations or frames.

\section{Results}
\label{sec:Results}

\begin{figure*}
  \centering
  \includegraphics[width=0.95\textwidth]{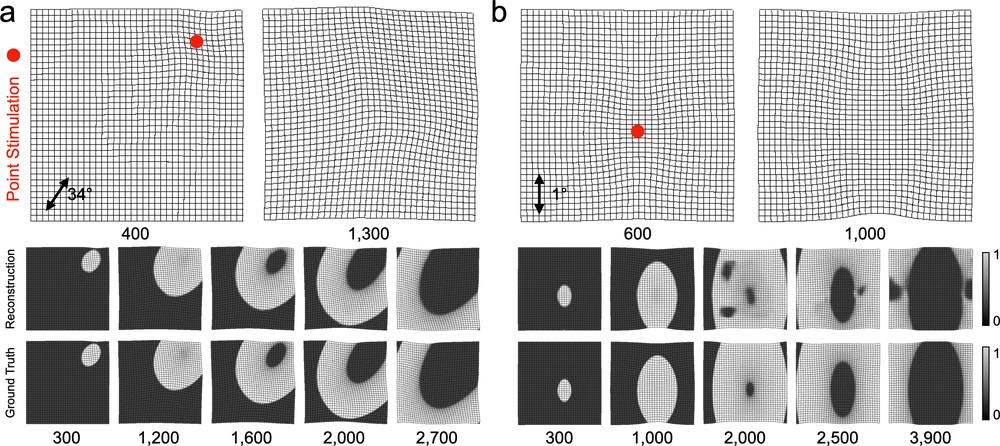}
  \caption{
    Reconstruction of focal electrical excitation wave patterns in two-dimensional deforming elastic excitable medium with fiber anisotropy using deep learning.
    The focal wave patterns originate from unknown, randomly chosen stimulation sites (red dots, at $t=0$).
    Top: mechanical deformation (displayed as mesh, cf. Fig. \ref{fig:kinematics}) analyzed by autoencoder neural network. The fiber direction is not known to the autoencoder.
    Bottom: reconstructed and original ground truth electrical excitation wave patterns $\widetilde{V}(x,y)$ and $V(x,y)$, respectively, which caused the deformation (normalized units [0,1], white: excited tissue, black: resting tissue, $t$ in simulation time steps).
    \textbf{a)} Stimulation site at $\vec{x}_S = (150,159)$ and $34^{\circ}$ fiber alignment. 
    \textbf{b)} Stimulation site at $\vec{x}_S = (110,84)$ and $1^{\circ}$ fiber alignment. 
    With focal data the autoencoder neural network (models 2Dt-A3f or 2Dt-A3, see table \ref{table:networkarchitecture} and section \ref{sec:results:trainingbias}) achieves in the large majority of cases very high reconstruction accuracies of about $99\%$ as shown in a).
    Due to (relatively rare) artifacts as shown in b) the average reconstruction accuracy is slightly lower, see also Supplementary Movie 7 for an impression of the reconstruction accuracy over time and across stimuli, and see also table \ref{table:networkarchitecture}.
  }
  \label{fig:results2Dfocal}
\end{figure*}

\begin{figure*}
\centering
\includegraphics[width=0.99\textwidth]{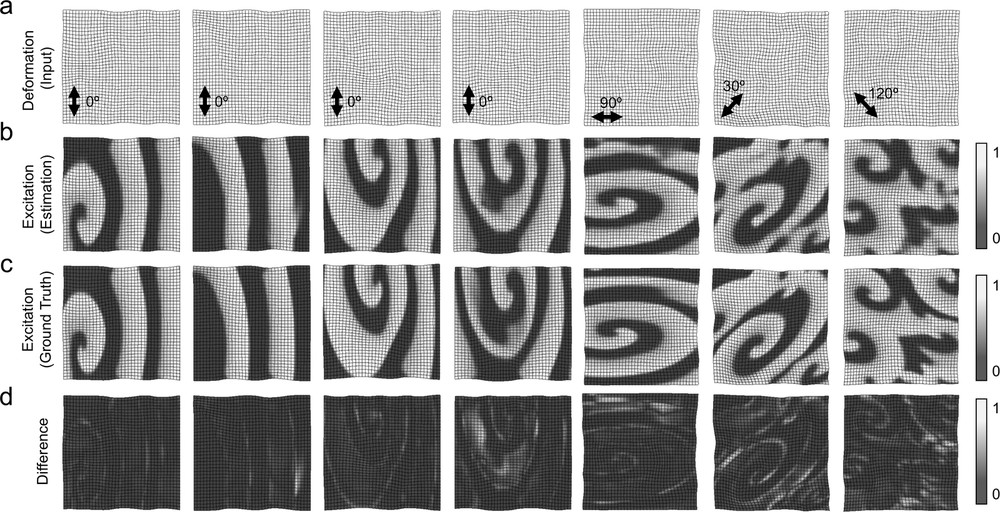}
\caption{
Reconstruction of electrical spiral and plane wave patterns from mechanical deformation using machine learning algorithm: 
\textbf{a)} Input data processed by autoencoder neural network: mechanical deformation data consisting of (time-varying) $u_x$- and $u_y$-displacements showing tissue contractions and deformations caused by electrical excitation wave patterns in c). Muscle fiber orientation indicated by black arrows.
\textbf{b)} Reconstructed electrical excitation wave pattern $\widetilde{V}(x,y)$ estimated by autoencoder from mechanical deformation shown in a) with an average reconstruction accuracy of about $95\%$, see table \ref{table:networkarchitecture}.
\textbf{c)} Ground truth excitation wave pattern $V(x,y)$ that caused the deformation shown in a).
\textbf{d)} Absolute error $\Delta V (x,y) = |\widetilde{V}(x,y)-V(x,y)|$ between estimated and ground truth excitation wave pattern.
The reconstructions shown in this figure were performed with model 2Dt-A3s, which was trained with chaotic spiral wave data.
}
\label{fig:results2D}
\end{figure*}

\begin{figure*}
  \centering
  \includegraphics[width=0.95\textwidth]{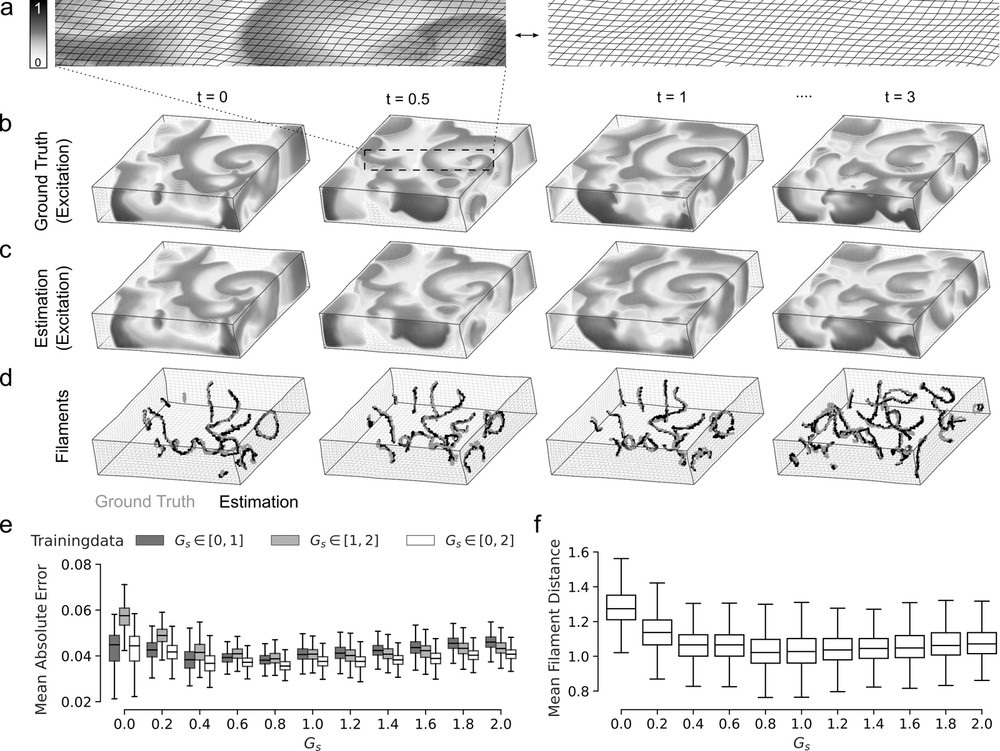}
  \caption{
    Reconstruction of three-dimensional electrical excitation wave dynamics from mechanical deformation (using model 3Ds-A1, see table \ref{table:networkarchitecture}).
    \textbf{a)} Close-up of deforming bulk surface. Left: electrical excitation (normalized units [0,1], gray: excited tissue, white/transparent: resting tissue). Right: mechanical deformation induced by excitation, which is analyzed by the autoencoder neural network. 
    To enhance viewing of the deformation, it's magnitude was exaggerated (multiplied by a factor of 2) in this panel. An impression of the original deformation is given in Supplementary Movie 3.
    \textbf{b)} Original electrical scroll wave pattern $V(x,y,z)$ generated with computer simulation.
    \textbf{c)} Reconstructed electrical scroll wave pattern $\widetilde{V}(x,y,z)$, which was estimated by neural network from deformation, and which is visually nearly indistinguishable from the original electrical scroll wave pattern.
    \textbf{d)} Comparison of original (gray) and reconstructed (black) electrical scroll wave vortex filaments, see also Supplementary Movie 2.
    \textbf{e)} Reconstruction error, given as average absolute difference between $\widetilde{V}$ and $V$ (in $\%$), over a wide range of mechano-electrical feedback strengths ($G_s=0,...,2$). The reconstruction performs robustly even if the network was trained only partially (grey) with $G_s \in [0,1]$ or $G_s \in [1,2]$. None of the data used for determining the error was included in the training.
    \textbf{f)} Mean distance between original and reconstructed electrical vortex filaments of $\sim 1.1 \pm 0.2$ voxel over a wide range of mechano-electrical feedback strengths ($G_s=0,...,2$).
  }
  \label{fig:results3D1}
\end{figure*}

Autoencoder neural networks can be used to compute electrical excitation wave patterns from mechanical motion and deformation in generic two-dimensional sheet- or three-dimensional bulk-shaped numerical simulations of cardiac muscle tissue with very high reconstruction accuracies of $90-98\%$, see table \ref{table:networkarchitecture}.
Various two- and three-dimensional electrical excitation wave phenomena, such as focal waves, planar waves, spiral and scroll waves and spatio-temporal chaos, can be reconstructed from mechanical motion and deformation, even in the presence of noise, low resolution and using either spatio-temporal or static mechanical data, see Figs. \ref{fig:results2Dfocal}-\ref{fig:noise} and Supplementary Movies 1-10.

\subsection{Recovery of Two-Dimensional Electrical Excitation Wave Patterns from Mechanical Deformation}\label{sec:results2D}

Figs. \ref{fig:results2Dfocal} and \ref{fig:results2D} demonstrate that various two-dimensional electrical excitation wave patterns, such as focal waves originating from random stimulation sites, as well as linear waves, single spiral and multi-spiral wave patterns, and even spiral wave chaos, can be reconstructed with very high precision from the resulting mechanical deformation using an autoencoder neural network, see also Supplementary Movies 1 and 7.
It is important to take notice of the fact that the neural network solely processes and analyzes tissue displacements, and does not analyze pre-computed strains or stresses during the processing, and does not have direct knowledge about the underlying cardiac muscle fiber alignment.
Neither during training nor during the reconstruction, was the underlying muscle fiber direction known to the network.
However, the contractile forces triggered by the electrical excitation wave patterns act along an underlying linearly transverse muscle fiber organization, which leads to highly anisotropic macroscopic contractions and deformations of the tissue, which is then reflected in the data.
The top panels in Fig. \ref{fig:results2Dfocal}a,b) and Fig. \ref{fig:results2D}a) show the deformations that were analyzed by the network, the two top rows in Fig. \ref{fig:results2Dfocal}a,b) and Fig. \ref{fig:results2D}b) show the reconstructed electrical excitation patterns $\tilde{u}(x,y)$, and the two bottom rows in Fig. \ref{fig:results2Dfocal}a,b) and Fig. \ref{fig:results2D}c) show the original ground truth excitation patterns $u(x,y)$ during focal activity and during spiral wave chaos, respectively. 
The different muscle fiber alignments are indicated by black arrows: $40^{\circ}$ in Fig. \ref{fig:results2Dfocal}a), $1^{\circ}$ in Fig. \ref{fig:results2Dfocal}b), and $0^{\circ}$ or linearly transverse in $e_y$-direction, $90^{\circ}$ or linearly transverse in $e_x$-direction, $30^{\circ}$ and $120^{\circ}$ from left to right in Fig. \ref{fig:results2D}a).

\begin{figure}
  \centering
  \includegraphics[width=0.48\textwidth]{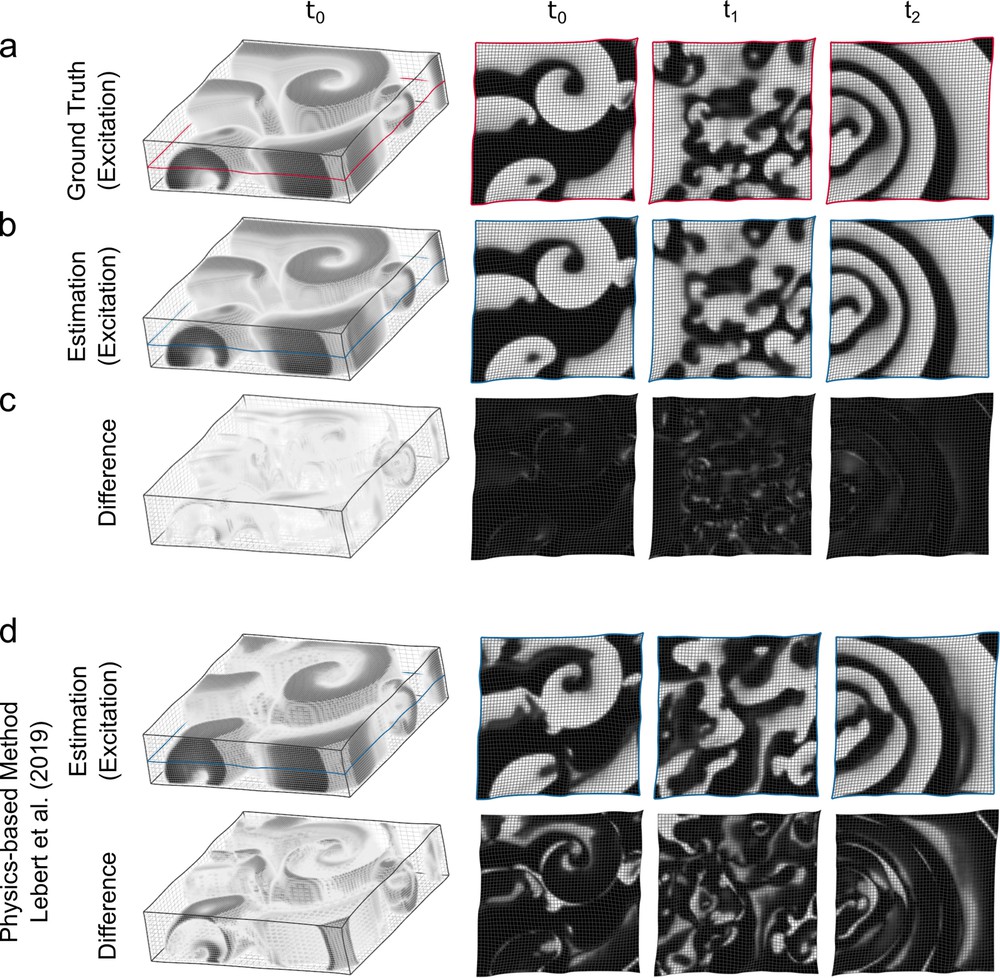}
  \caption{Reconstruction accuracy of data-driven (model 3Ds-A1) compared to physics-based (Lebert et al. \cite{Lebert2019}) mechano-electrical reconstruction approach. 
  \textbf{a-d)} Volume renderings (gray: depolarized, transparent: resting) and cross-sections (white: depolarized, black: resting) showing \textbf{a)} original electrical excitation $V(x,y,z)$, \textbf{b)} electrical excitation $\widetilde{V}(x,y,z)$ reconstructed using data-driven approach, \textbf{c)} difference $\Delta V = |\widetilde{V}-V|$, and \textbf{d)} electrical excitation $\widetilde{V}^*(x,y,z)$ reconstructed using physics-based approach and difference to original $\Delta V = |\widetilde{V}^* - V|$, respectively, at time steps $t_0$, $t_1$, $t_2$. The data-driven approach outperforms the physics-based approach.
  }
  \label{fig:results3D2}
\end{figure}

\begin{figure*}
  \centering
  \includegraphics[width=0.98\textwidth]{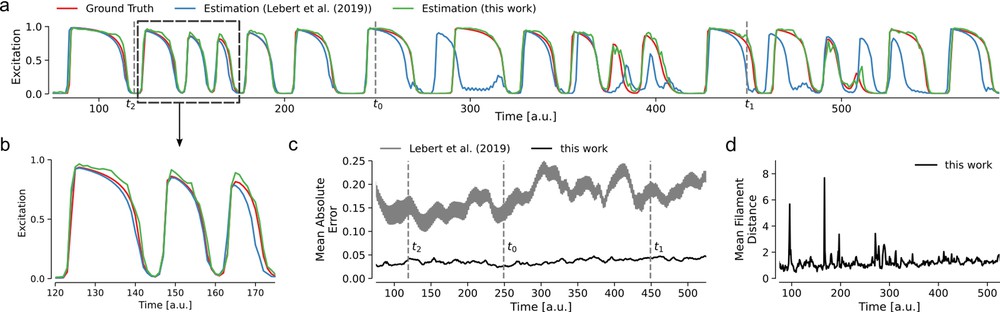}
  \caption{Reconstruction accuracies obtained with machine learning (model 3Ds-A1) and physics-based approach \cite{Lebert2019} for three-dimensional electromechanical fibrillation or scroll wave  dynamics.
  \textbf{a)} Exemplary time-series of ground truth excitation $V(t)$ (red), reconstructed excitation $\widetilde{V}(t)$ obtained with machine learning approach (green), and reconstructed excitation $\widetilde{V}^*(t)$ obtained with physics-based approach (blue).
  \textbf{b)} Close-up of a) during wave break-up at $t_2$, cf. Fig. \ref{fig:results3D2}.
  \textbf{c)} Reconstruction error over time given as mean absolute error $\langle \Delta V \rangle = \sum |V-\widetilde{V}|/N$. Reconstruction error remains small ($<5\%$) with machine learning approach for all regimes: planar waves at time $t_0$, scroll waves at time $t_1$, and scroll wave chaos at time $t_2$, cf. Fig. \ref{fig:results3D2}.
  \textbf{d)} The average filament distance between original (ground truth) and reconstructed scroll wave filaments obtained with machine learning approach in the timespan shown is $1.2\pm0.5$ voxels and stays very small in all regimes. The occasional peaks arise when a new filament does not have a corresponding filament in the estimation. The average filament distance can not be calculated over time with physics-based approach due to insufficient quality of the reconstruction.
  }
  \label{fig:results3D3}
\end{figure*}

In many cases, the autoencoder's reconstructions are almost visually indistinguishable from the original excitation patterns, e.g. as seen in Fig. \ref{fig:results2Dfocal}a) or \ref{fig:results2D}, with reconstruction errors $\langle \Delta V \rangle$ in the order of $2\%-6\%$.
On spiral chaos data (model 2Dt-A3s, see table \ref{table:networkarchitecture}) the reconstruction accuracy is $95.1 \% \pm 2.6 \% (\pm 8.5 \%)$, on focal data (model 2Dt-A3f) the reconstruction accuracy is $98.6 \% \pm 1.4 \% (\pm 4.9 \%)$, and on a mix of spiral chaos and focal data (50\%-50\% mix, model 2Dt-A3) the reconstruction accuracy is $96.2 \% \pm 3.1 \% (\pm 7.8 \%)$, respectively, see Figs. \ref{fig:results2Dfocal} and \ref{fig:results2D} and also sections \ref{sec:results:temporal}-\ref{sec:results:trainingbias} for a more detailed discussion.
Fig. \ref{fig:results2D}d) shows the absolute difference between reconstructed and ground truth excitation $\Delta V = |\widetilde{V}(x,y)-V(x,y)|$ during spiral wave chaos. One sees that residual reconstruction errors occur mostly at the waves fronts and backs, but that the reconstructed and original excitation patterns are largely congruent. In Fig. \ref{fig:results2Dfocal}b) one notices that the reconstructions at times also contain clearly visible artifacts, particularly for focal waves with longer wavelengths. However, the larger artifacts are relatively rare, see also Supplementary Movie 7 for an impression of their prevalence over time and across stimuli.
Accordingly, the standard deviation $\sigma_{\langle V \rangle}$ across frames is in the order of $\pm 2\%$ to $\pm5\%$, whereas the standard deviation $\sigma_V$ across all pixels is in the order of $\pm 5\%$ to $\pm10\%$ for both focal and spiral wave chaos data.
Overall, the inverse mechano-electrical reconstruction's accuracy is very high and consistently greater than $90\%$, regardless of the particular neural network model used, see table \ref{table:networkarchitecture} and sections \ref{sec:results:temporal}-\ref{sec:results:resolution}. 
Even spiral waves with both long and short and changing wavelengths during breathing instabilities or alternans can be recovered from analyzing the correspondingly induced mechanical deformation, see third and fourth column in Fig. \ref{fig:results2D}.
During spiral wave chaos the reconstruction error does not change significantly over time, regardless of whether single spiral waves, multiple spiral waves or spiral wave chaos are estimated with the same autoencoder network (e.g. model 2Dt-A3s), see also Fig. \ref{fig:results3D2}c).
Training the network models on either just focal data or just spiral chaos data, as shown in Fig. \ref{fig:trainingdata}a) or b), or on a mixture of both data types may be advantageous, see section \ref{sec:results:trainingbias} for a more detailed discussion on training bias.
The estimation of a two-dimensional excitation wave pattern ($128 \stimes 128$ pixels) can be performed within less than $0.8\,\text{ms}$ by the neural network (model 2Dt-A3, average from $1,000$ frames, Nvidia GPU, see methods).
The data shown in Figs. \ref{fig:results2Dfocal}-\ref{fig:results2D} and \ref{fig:temporal}-\ref{fig:noise} was not seen by the network during the training procedure.

\subsection{Recovery of Three-Dimensional Electrical Excitation Wave Patterns from Mechanical Deformation}\label{sec:results3D}

Figs. \ref{fig:results3D1}-\ref{fig:results3D2} demonstrate that various three-dimensional electrical excitation wave patterns, such as planar or spherical waves, single scroll waves or even scroll wave chaos, can be reconstructed from the resulting mechanical deformation using an autoencoder neural network with three-dimensional convolutional layers, see also Supplementary Movies 2-4.
The neural network solely processes and analyzes three-dimensional tissue motion, see Fig. \ref{fig:results3D1}a). The deformation shown in panel a) shows the bulk surface, but three- dimensional kinematic data is used for the reconstruction (the displayed deformation is exaggerated to facilitate viewing of length changes, displacements were multiplied by factor $2$). Supplementary Movie 3 shows the corresponding deformation over time (original without exaggerating the motion's magnitude).
As in the two-dimensional tissue, the contractile forces act along muscle fibers, which rotate throughout the three-dimensional bulk by a total angle of $120^{\circ}$, see section \ref{sec:methods:simulations}.
Fig. \ref{fig:results3D1}b) shows volume renderings of the original three-dimensional electrical excitation wave dynamics $V(x,y,z)$ that led to the deformations of the bulk. 
These deformations were then analyzed by the autoencoder network. The electrical wave dynamics correspond to scroll wave chaos that originated from spherical and planar waves through a cascade of wave breaks. 
Panel c) shows volume renderings of the reconstructed electrical excitation wave pattern $\widetilde{V}(x,y,z)$ that was estimated from the bulk's deformations.
The reconstructed electrical scroll wave pattern is visually indistinguishable from the original electrical scroll wave pattern. 
The reconstruction error $\langle \Delta V \rangle$ is in the order of $4 \% \pm 1 \%$, similar as seen with the two-dimensional reconstructions in section \ref{sec:results2D}.
The reconstruction error remains small for a broad range of electrical parameters, e.g. for mechano-electrical feedback strengths ranging from $G_s=0.0,...,2.0$, see Fig. \ref{fig:results3D1}e), even if the network was trained only partially on $G_s$ (gray: $G_s \in [0,1]$ or $G_s \in [1,2]$, white: $G_s$ from whole range $[0,2]$ included in training, error largely unaffected).
None of the data used for reconstruction and determining the error was used during training, see also section \ref{sec:methodstraining}.
The evolution of the reconstructed electrical scroll wave dynamics appears smooth, see also Supplementary Movie 2, even though each three-dimensional volume frame was estimated or reconstructed individually by the neural network, cf. section \ref{sec:results:temporal}.
Individual time-series show that the action potential shapes and upstrokes of the electrical excitation are reconstructed robustly with minor deviations from the ground truth, see Fig. \ref{fig:results3D3}a,b).
The reconstruction error does not fluctuate much over time ($\sigma_{\langle V \rangle} < 1 \%$), even though the sequence contains planar waves, single scroll waves, as well as fully turbulent scroll wave chaos, see Fig. \ref{fig:results3D2}c,d) and Fig. \ref{fig:results3D3}c).
Fig. \ref{fig:results3D2} shows a comparison of the reconstruction accuracy obtained with the autoencoder with that obtained with a physics-based mechano-electrical reconstruction approach, which we published recently \cite{Lebert2019}. The difference maps in panel d) show substantially larger residual errors than with the autoencoder in panel c). Correspondingly, the reconstruction error shown in Fig. \ref{fig:results3D3}c) is more than 3-fold larger and fluctuates much more with the physics-based approach than with the autoencoder used in this work.
The estimation of a single three-dimensional excitation wave pattern ($100 \stimes 100 \stimes 24$ voxels) can be performed within $20 \text{ms}$ by the neural network (model 3Ds-A1, Nvidia GPU, see methods).
The data shown in Figs. \ref{fig:results3D1}-\ref{fig:results3D2} was not seen by the network during the training procedure.

\subsubsection{Recovery of Electrical Vortex Filaments from 3D Mechanical Deformation Data}

We computed electrical vortex filaments, or three-dimensional electrical phase singularities, which describe the cores or rotational centers of electrical scroll waves, from both the original and reconstructed excitation wave patterns via the Hilbert transform, see panel d) and Supplementary Movie 2.
Both filament structures are almost identical with an average distance between original and reconstructed filaments of $1.1 \pm 0.2$ voxel, see Fig \ref{fig:results3D1}f). The mismatch is in the order of the precision with which the spatial locations of the filaments can be computed.
One voxel corresponds to about $1\%$ of the medium size ($100 \stimes 100 \stimes 24$ cells). 
Except for a few short-term fluctuations, the mean distance between original and reconstructed electrical vortex filaments does not decrease or increase much over time, see Fig. \ref{fig:results3D3}d), and stays small for different dynamics and over a broad range of mechano-electrical feedback strengths $G_s=0.0,...,2.0$, see Fig. \ref{fig:results3D1}f).
The filament data shows that the topology of the electrical excitation wave pattern can be reconstructed with very high accuracy from deformation.

\subsection{Reconstruction from Spatio-Temporal Mechanical Deformation Data and with Arbitrary Reference Frames}
\label{sec:results:temporal}

\begin{figure}
\centering
\includegraphics[width=0.46\textwidth]{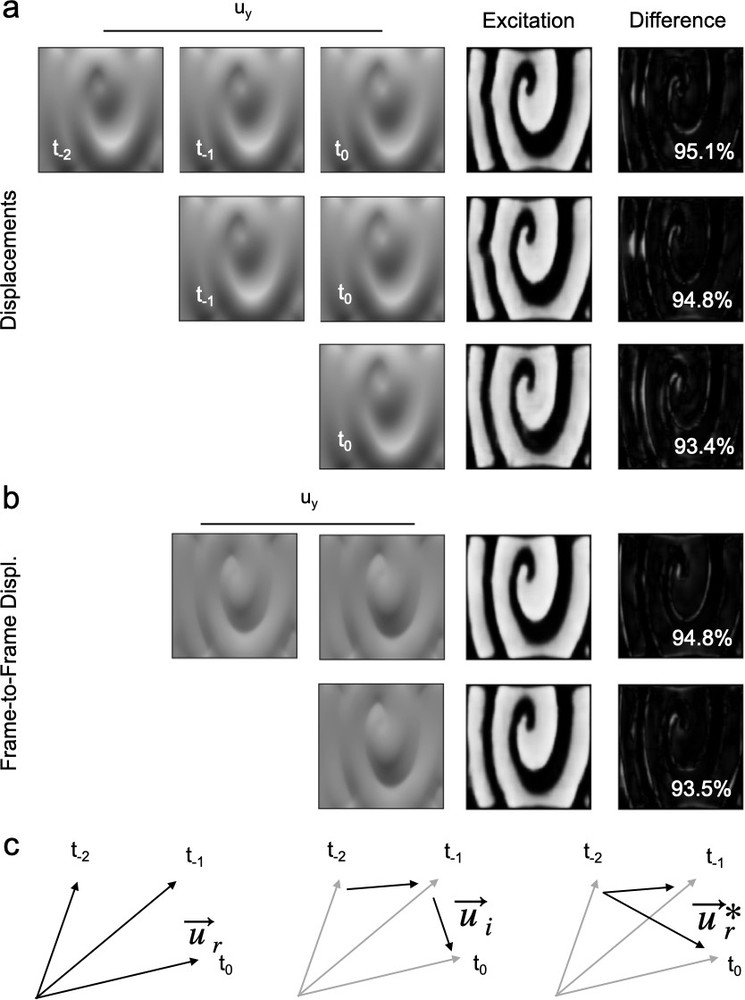}
\caption{
Learning from a temporal sequence of mechanical input patterns and different types of kinematic data.
Training and reconstruction accuracies with 
\textbf{a)} one, two or three consecutive mechanical frames (models 2Ds-A1s, 2Dt-A2s, 2Dt-A3s, respectively) with displacement vectors $\vec{u}_r$ (only $u_y$-component shown, scale: [-4, 4] pixels), or
\textbf{b)} with one or two consecutive mechanical frames (models 2Dt'-A1s and 2Dt'-A2s, respectively) with instantaneous frame-to-frame displacement vectors $\vec{u}_i$ (only $u_y$-component shown, scale: [-2, 2] pixels).
Right: reconstructed $\widetilde{V}(x,y)$ and difference to ground truth excitation wave pattern $\Delta V(x,y)$ and reconstruction accuracies $1-\langle \Delta V \rangle$ in percent [\%].
The mechanical frames have a temporal offset of $50$ simulation time steps ($\sim 5\%$ of spiral rotation period).
\textbf{c)} Different types of displacement data ($\vec{u}_r$, $\vec{u}_i$, $\vec{u}_r^*$) used for reconstruction.
}
\label{fig:temporal}
\end{figure}

Both static and short spatio-temporal sequences of mechanical deformation patterns can be processed by the autoencoder, depending on the number of input layers, see Fig. \ref{fig:autoencoder}b-c), and both lead to accurate reconstructions of the underlying excitation wave patterns, see Fig. \ref{fig:temporal}.
Panel a) shows that the reconstruction achieves an accuracy of $93.4\% \pm 3.2 \% (\pm 11.0 \%)$
when processing a single mechanical frame (using model 2Ds-A1s with 2 input channels for $u_x$- and $u_y$-components of displacement fields, respectively), which contains displacements $\vec{u}_r$ that describe the motion of each of the tissue's material segment with respect to their original position in the stress-free, undeformed reference configuration $\chi_0$ (only $u_y$-component shown with scale [-4, 4] pixels), see also schematic (left) in Fig. \ref{fig:temporal}c).
The accuracy can be further improved to $94.8\% \pm 2.6 \% (\pm 8.7 \%)$ 
and $95.1\% \pm 2.6 \% (\pm 8.5 \%)$ %
when feeding $2$ or $3$ consecutive mechanical frames (or $2$ or $3$ consecutive sets of $\vec{u}_r = \{ \vec{u}_r(t_0), \vec{u}_r(t_{-1}),...\}$) as input into the network (models 2Dt-A2s with $4$ or 2Dt-A3s with $6$ input channels), respectively, see also Supplementary Movie 5 and Fig. \ref{fig:temporal}c). 
Analyzing spatio-temporal mechanical data improves the reconstruction's accuracy and robustness (cf. models 2Ds-B1s and 2Dt-B3s in table \ref{table:networkarchitecture}), particularly also with low resolution mechanical data (cf. models 2Ds-F1s and 2Dt-F3s in section \ref{sec:results:resolution}.

Next, the autoencoder can also succesfully process and obtain highly accurate reconstructions with instantaneous displacement data $\vec{u}_i$, which describes the motion of the tissue from the previous time step(s) to the current time step (temporal offset $\sim 5\%$ of spiral period), see also Fig. \ref{fig:kinematics}c).
Fig. \ref{fig:temporal}b) shows that the reconstruction (model 2Dt'-A1s) achieves an accuracy of $93.5\% \pm 3.2 \% (\pm 11.0 \%)$
when processing a single mechanical frame, which contains instantaneous frame-to-frame displacements (only $u_y$-component shown with scale [-2, 2] pixels) calculated as the difference between two subsequent displacement vectors $\vec{u}_i(t_0) = \vec{u}_r(t_0) - \vec{u}_r(t_{-1})$ in time, see schematic (center) in Fig. \ref{fig:temporal}c). 
The accuracy increases slightly further to $94.8\% \pm 2.6\% (\pm 9.2 \%)$
when processing $2$ consecutive mechanical frames (model 2Dt'-A2s) containing instantaneous frame-to-frame displacements $\vec{u}_i = \{ \vec{u}_i(t_0),\vec{u}_i(t_{-1})\} = \{ \vec{u}_r(t_0) - \vec{u}_r(t_{-1}), \vec{u}_r(t_{-1})-\vec{u}_r(t_{-2})\}$, respectively, which were calculated as the difference between subsequent displacement vectors in time, see schematic (center) in Fig. \ref{fig:temporal}c). Analyzing displacements $\vec{u}_r$ or instantaneous frame-to-frame displacements $\vec{u}_i$ yields almost equal reconstruction accuracies.

Note that in some imaging situations, particularly during arrhythmias, the tissue's stress-free undeformed mechanical configuration $\chi_0$ may not be known, and accordingly displacement data $\vec{u}_r$ would not be readily available.
Nevertheless, the autoencoder neural network can also correctly reconstruct excitation wave patterns, even if the input to the network are displacement vectors $\vec{u}_r^*$ that indicate motion and deformation of the tissue segments with respect to an arbitrary deformed configuration $\chi_t$ in an arbitrary frame, see schematic (right) in Fig. \ref{fig:temporal}c). 
Analyzing the tissue's relative motion with respect to an arbitrary deformed configuration $\chi_t$, the inverse mechano-electrical reconstruction nevertheless achieves reconstruction accuracies of $93.5\% \pm 3.0 \% (\pm 10.7 \%)$ and $94.8\% \pm 2.6 \% (\pm 9.1 \%)$ when analyzing $1$ (model 2Ds$^*$-A1s) or $2$ (model 2Dt$^*$-A2s) mechanical frames, respectively, each displacement vector indicating shifts of the tissue $\vec{u}_r^*$ relative to a deformed configuration $\chi_{t-2}$ a few time steps prior to the current frame. In more detail, in a series of $3$ subsequent mechanical frames the current ($t_0$) and/or previous frame ($t_{-1}$), which indicate motion with respect to the tissue's configuration in the first frame of that series ($t_{-2}$), was/were analyzed by the network. The network models were trained accordingly.

The data demonstrates that the inverse mechano-electrical reconstruction can be performed robustly with both static or time-varying mechanical input data, as well as with absolute or instantaneous frame-to-frame displacement data, which can alternatively describe motion with respect to the undeformed, stress-free or an arbitrary reference configuration. 
Principally, similar kinematic data can be retrieved with numerical motion tracking in imaging experiments.

\subsection{Effect of Training Duration, Training Dataset Size and Neural Network Size onto Reconstruction Accuracy}
\label{sec:results:training}

\begin{figure}
\centering
\includegraphics[width=0.46\textwidth]{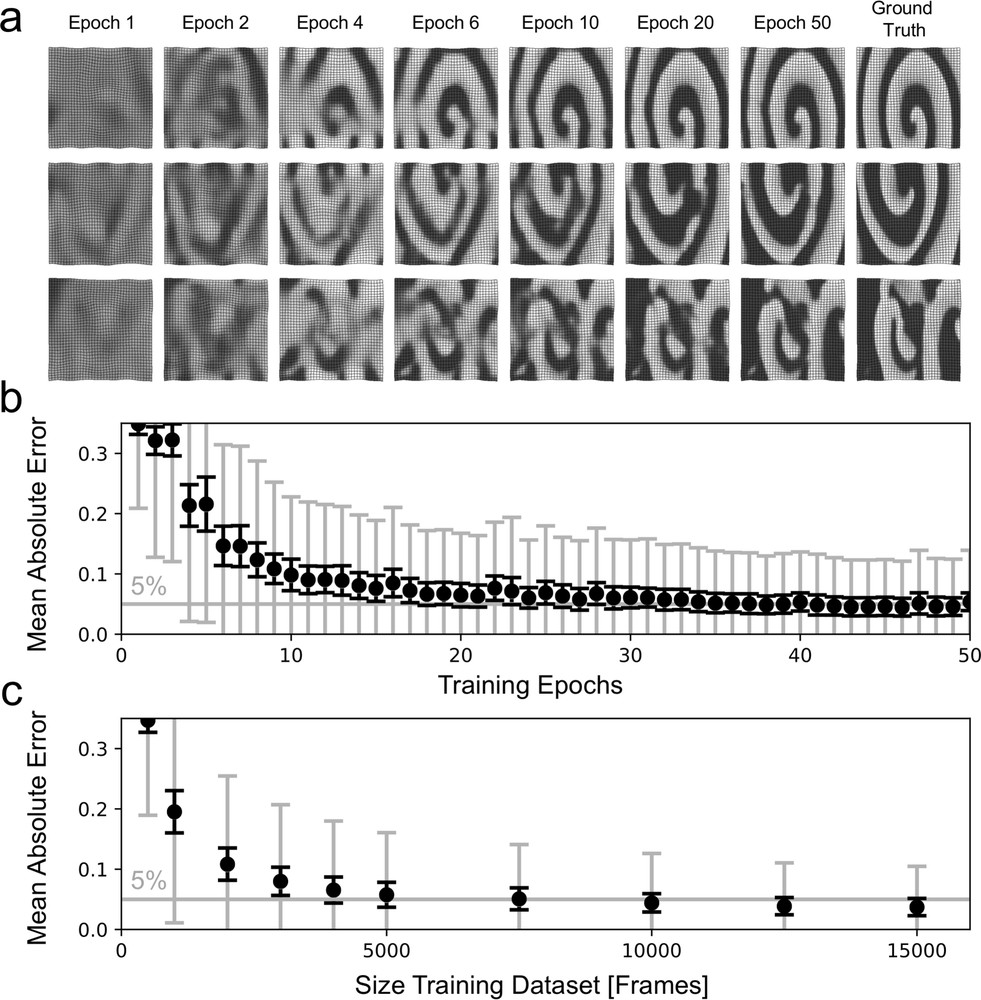}
\caption{
Reconstruction accuracy (model 2Dt-A3s) improves with training duration and size of training dataset.
\textbf{a)} Examplary reconstructions for training epochs $n_e = 1,2,4,6,10,20,50$ (or increasing training duration) and ground truth (right). Quality of reconstruction increases with training epochs.
\textbf{b)} Reconstruction error $\langle \Delta V \rangle$ over number of training epochs.
\textbf{c)} Reconstruction error $\langle \Delta V \rangle$ over size of training dataset in number of frames.
Black error bars: standard deviation $\sigma_{\langle V \rangle}$ across frames. Gray error bars: standard deviation $\sigma_V$ across all absolute difference values $|\widetilde{V}-V|$.
}
\label{fig:training}
\end{figure}

The training duration and the size of the training dataset are generally important parameters in data-driven modeling.
Fig. \ref{fig:training} shows how the training duration, measured in training epochs, and the size of the training dataset affect the reconstruction error for two-dimensional data (model 2Dt-A3s, see table \ref{table:networkarchitecture}).
The image series in Fig. \ref{fig:training}a) demonstrate how the reconstruction accuracy increases with training duration. The excitation wave pattern can already be identified after $2-5$ epochs, however, with substantial distortions. 
After $10-20$ epochs, the reconstruction progressively recovers and resolves finer details, and after $40-50$ epochs also residual artifacts are removed to a large extent, see also Supplementary Movie 6.
The ground truth excitation wave pattern $V(x,y)$ is shown on the right.
Accordingly, panel b) shows the mean absolute error, see eq. (\ref{eq:error}), plotted over the number of training epochs.
The reconstruction error $\langle \Delta V \rangle$ quickly decreases to below $10\%$ just after a few training epochs, and approaches a mean absolute error of about $5\%-10\%$ after $10-20$ training epochs. The error continuously decreases and saturates at about $4\%-5\%$ for $40-50$ epochs. After $200$ training epochs the reconstruction accuracy (model 2Dt-A3s) is $96.8\% \pm 1.6\% (\pm 5.7 \%)$ compared to $95.1\% \pm 2.6\% (\pm 8.5\%)$ after $50$ epochs, respectively.
All reconstructions in this study were obtained with $50$ training epochs if not stated otherwise.
Note that the (gray) error bars in Fig. \ref{fig:training}b,c) reflect the larger residual errors.
Overall, the reconstruction error $\langle \Delta V \rangle$ is comparable in two- and three-dimensional tissues, cf. Figs. \ref{fig:results3D1}e) and \ref{fig:results3D3}c). %
Panel c) demonstrates how the size of the training dataset determines the reconstruction accuracy.
Generally, the larger the training dataset, the better the reconstruction. However, the reconstruction accuracy does not appear to improve significantly with more than $20,000$ frames.
In our study, the number of network model parameters affected the network's performance only slightly. For instance, smaller two-dimensional models with about $300,000$ trainable model parameters achieve a slightly lower reconstruction accuracy (e.g. model 2Dt-B3s: $94.5 \% \pm 2.7 \%$) than larger models with more than $1,000,000$ model parameters on the same data (e.g. model 2Dt-A3s: $95.1 \% \pm 2.6\%$).

\subsection{Training Bias}
\label{sec:results:trainingbias}
We found that the training data can influence the model’s reconstruction accuracy or generate a bias towards the training data.
With the two-dimensional neural network models, the reconstruction error worsened substantially if the training was performed, for instance, only with focal wave patterns, and then, subsequently, the reconstruction was applied on spiral wave patterns.
In the following, we distinguish different network models based on the type of data they were trained on, focal (f) and/or spiral chaos (s) data, see also table \ref{table:networkarchitecture}, and assess their reconstruction errors.
If the network (model 2Dt-A3) was trained on both focal and spiral wave chaos data (50\%/50\% ratio in $20,000$ samples), its reconstruction error for a 50\%-50\%-mix of both focal and spiral wave chaos data is $3.8 \% \pm 3.1 \% (\pm 7.8 \%)$, for focal wave data it is $1.6 \% \pm 1.3 \%$ ($\pm 4.4 \%$), and for spiral wave chaos data it is $6.0 \% \pm 2.8 \% (\pm9.9\%)$, respectively. 
However, if the network (model 2Dt-A3f) was trained just on focal (f) data, it performs very well on focal data, as expected, its reconstruction error being $1.4 \% \pm 1.4 \% (4.9 \%)$, but poorly on spiral chaos data, on which the reconstruction error becomes $20.7\% \pm 6.9 \% (\pm 25.7\%)$, respectively. Accordingly, the same network yields a mediocre overall reconstruction error of $11.0\% \pm 10.8\% (\pm 20.9 \%)$ with a 50\%-50\%-mix of both focal and spiral wave chaos patterns, where the error largely results from analyzing the spiral wave patterns, while the focal patterns are reconstructed accurately.
Vice versa, if the network (model 2Dt-A3s) was trained just on spiral chaos data, the network's reconstruction error is
$4.9\% \pm 2.6\% (\pm 8.5\%)$
for spiral chaos data, 
$11.3\% \pm 7.8 \% (\pm 15.5\%)$ for focal data, and
$5.8\% \pm 6.2 \%(\pm 12.3 \%)$ 
for a 50\%-50\%-mix of both patterns, respectively.
The results suggest, firstly, that the careful selection of electrical and mechanical training data will be critical in future imaging applications, and, secondly, that specialized neural networks trained exclusively on particular types of rhythms or arrhythmias may achieve higher reconstruction accuracies in specialized applications than general networks (cf. model 2Dt-A3f vs. 2Dt-A3 on focal data). 
However, in some situations, the increase in accuracy associated with specialization might be negligible, as general networks trained on various rhythms may already achieve similarly high reconstruction accuracies as well (c.f. model 2Dt-A3 vs. 2Dt-A3s on spiral wave chaos data).

\subsection{Robustness against Measurement Noise}

\begin{figure}
\centering
\includegraphics[width=0.46\textwidth]{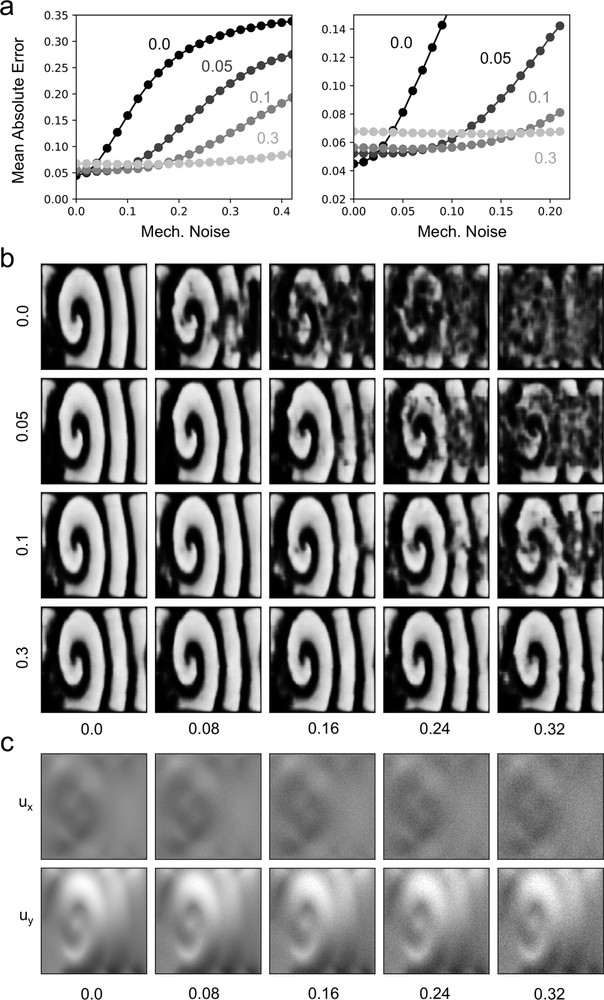}
\caption{
Reconstruction accuracy (model 2Dt-A3s) with noisy mechanical input data.
\textbf{a)} Reconstruction error $\langle \Delta V \rangle$ increases with noise, but can be kept minimal when models are trained with noisy training data (different curves with $\tilde{\xi} = 0.0, 0.05, 0.1, 0.3$ noise in training data). Right, close-up: training with noise deteriorates baseline accuracy (at $\xi=0.0$).
\textbf{b)} Reconstructed excitation patterns $\widetilde{V}(x,y)$ in the presence of noise. Vertical: amount of noise $\tilde{\xi}$ during training ($0.0$, $0.05$, $0.1$, $0.3$). Horizontal: amount of noise $\xi$ during estimation ($0.0$, $0.08$, $0.16$, $0.24$, $0.32$). 
\textbf{c)} Displacement components $u_x$ and $u_y$ with increasing noise $\xi$.
}
\label{fig:noise}
\end{figure}

Mechanical measurement data obtained in imaging experiments is likely to contain noise (from noisy image data or intrinsic noise of numerical tracking algorithms).
To strengthen the network's ability to reconstruct excitation wave patterns with noisy mechanical input data, we added Gaussian white noise to the mechanical training data that otherwise contained smooth motion or displacement vector fields produced by the computer simulations.
Since denoising is one of the particular strengths of autoencoders \cite{Vincent2008,Gondara2016}, they should be particularly suited to handle noise as they perform mechano-electrical reconstructions.
Indeed, Fig. \ref{fig:noise} shows that if the two-dimensional network models are trained with noisy mechanical input data, they develop the capability to estimate the excitation despite the presence of noise in the mechanical input data.
The data also shows that, on the other hand, when presenting noisy mechanical input data to a network that was not trained with noise, the reconstruction accuracy quickly deteriorates. 
Fig. \ref{fig:noise}a) shows four curves representing the reconstruction errors $\langle \Delta V \rangle$ that were obtained with increasing mechanical noise $\xi$ with four different network models, one being trained without noise (black: $\tilde{\xi}=0.0$) and three being trained with noise (gray: $\tilde{\xi} = 0.05, 0.1, 0.3$), respectively.
The noise is normally distributed and was added onto the $u_x$- and $u_y$-components of the displacement vectors, see panel c). All stated values correspond to the standard deviation $\sigma$ of the noise. Noise with a magnitude of $\xi \approx 0.1$ pixels corresponds to approximately $10\%$ of the distance between displacement vectors. 
Without noisy training, the reconstruction error increases steeply if noise is added during the reconstruction procedure, almost two-fold for small to moderate noise levels of $\xi \approx 0.05$.
In contrast, training with noise (dark gray curves: $\tilde{\xi} = 0.05,0.1,0.3$) flattens the error curve substantially and yields acceptable reconstruction errors ($\langle \Delta V \rangle < 6\%$) up to the noise levels that were used during training.
However, training with noise comes at the cost of increasing the error at baseline (light gray curve: $\langle \Delta V \rangle > 6\%$ with $\tilde{\xi} = 0.3$).
Fig. \ref{fig:noise}b) shows the corresponding reconstructions with different noise levels $\xi = 0.0, 0.08, 0.16, 0.24, 0.32$ during estimation (horizontal), for the four different models with training noise $\tilde{\xi} = 0.0, 0.05, 0.1, 0.3$ (vertical).
Based on the results in Fig. \ref{fig:noise} we used training data that contained both noisy and noise-free mechanical training data.

\subsection{Lower Spatial Resolution of Mechanical Data} 
\label{sec:results:resolution}
\begin{figure}
\centering
\includegraphics[width=0.46\textwidth]{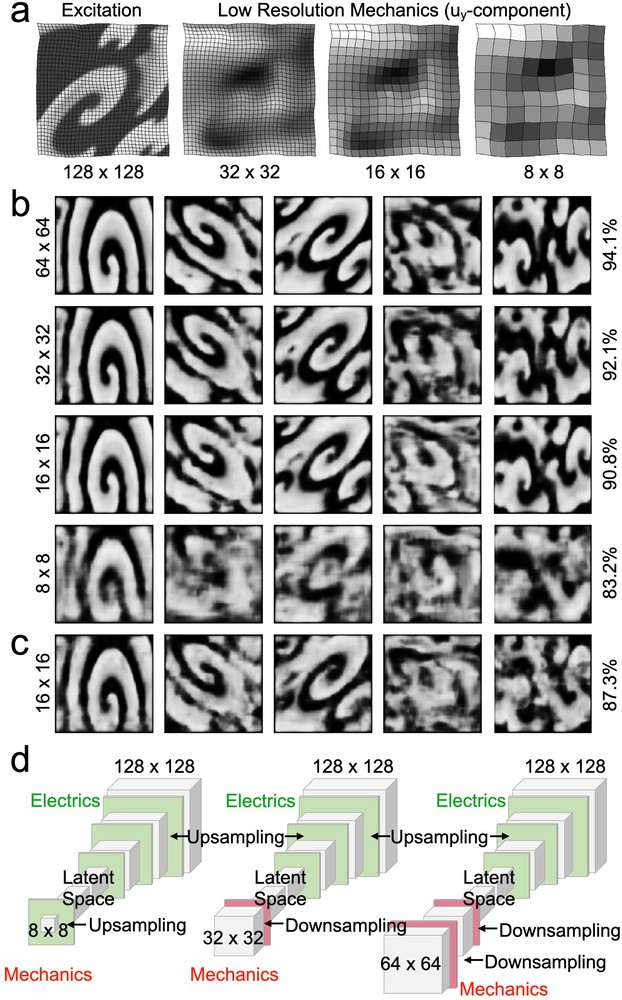}
\caption{
Reconstruction of electrical excitation patterns (all $128 \stimes 128$ pixels) from lower resolution mechanical deformation patterns ($64 \stimes 64$, $32 \stimes 32$, $16 \stimes 16$, $8 \stimes 8$ vectors).
\textbf{a)} Left: Excitation pattern at original resolution ($128 \stimes 128$). Right: Low resolution (sparsely sampled) versions of the same mechanical pattern (displayed as $u_y$-component of displacement vector plotted on deformed grid, cf. Fig. \ref{fig:kinematics}d)).
\textbf{b)} Degrading quality of excitation patterns (all $128 \stimes 128$ pixels) reconstructed from $2\stimes-16\stimes$ smaller mechanical deformation patterns (top to bottom, sparsely sampled). Sufficiently high reconstruction accuracies with up to $10\stimes$ smaller mechanical deformation than electrical patterns, see also Supplementary Movie 9.
\textbf{c)} Reconstructions as in b), but obtained with static low resolution ($8\stimes$) mechanical data.
\textbf{d)} Different truncated autoencoder network architectures with smaller input layer sizes ($8 \stimes 8$, $16 \stimes 16$, $32 \stimes 32$) than output layer sizes (here always $128 \stimes 128$) and constant latent space size ($16 \stimes 16$).
}
\label{fig:resolution}
\end{figure}

Mechanical measurement data obtained in imaging experiments may retain a lower spatial resolution than desired. 
Since enhancing the resolution of image data (superresolution) is a particular strength of autoencoders\cite{Zeng2017}, they should be particularly suited to perform inverse mechano-electrical reconstructions also with sparse or low resolution mechanical data.
Fig. \ref{fig:resolution} demonstrates that our autoencoder network is able to provide sufficiently accurate reconstructions, even if the mechanical input data has a lower spatial resolution, and illustrates the degree of deterioration of the reconstruction accuracy with increasingly lower spatial resolutions.
The original size of the two-dimensional electrical or mechanical data used throughout this study for training and reconstructions is $128 \stimes 128$ pixels.
To emulate lower spatial resolutions of the mechanical data, we downsized it to $64 \stimes 64$, $32 \stimes 32$, $16 \stimes 16$ and $8 \stimes 8$ displacement vectors in each time step, e.g. by repeatedly subsampling or leaving out every second displacement vector, respectively, and adjusted the input layers of the autoencoder accordingly, while keeping the output layer for the electrics constant at $128 \stimes 128$ pixels. The downsized mechanical displacement data was scaled accordingly (e.g. factor $0.5$ with $64 \stimes 64$ pixels).
The number of convolutional and downsampling (maxpooling) or upsampling layers in the encoding and decoding parts of the original network (model 2Dt-A3s) were modified to achieve the desired upsampling of lower resolved mechanical data to excitation data with size $128 \stimes 128$ pixels, see Fig. \ref{fig:resolution}d). For instance, mechanical data with size $64 \stimes 64$ was read by a network (model 2Dt-D3s) with an input layer of size $64 \stimes 64 \stimes 6$, and encoded by two stages with two $32 \stimes 32 \stimes 128$ and $16 \stimes 16 \stimes 256$ convolutional layers and two maxpooling layers in between, see table \ref{table:networkarchitecture}.
Mechanical data with size $16 \stimes 16$ was provided directly after a $16 \stimes 16 \stimes 64$ convolutional layer to the latent space (e.g. models 2Dt-Fxx), and data with size $8 \stimes 8$ was first upsampled and then send through a $16 \stimes 16 \stimes 64$ convolutional layer before being provided to the latent space, see Fig. \ref{fig:resolution}d).
Fig. \ref{fig:resolution}b) shows the reconstructions, all $128 \stimes 128$ pixels in size, of various electrical excitation wave patterns for different mechanical input sizes ($64 \stimes 64$ to $4 \stimes 4$ vectors, or subsampling / downsizing factor of $2\stimes-32\stimes$) provided to the autoencoder network. 
Remarkably, the autoencoder achieved reconstruction accuracies of $94.1\% \pm 3.1 \% (\pm 10.2 \%)$,  %
$92.1\% \pm 3.7 \% (\pm 11.8 \%)$, 
$90.8\% \pm 4.7 \% (\pm 13.6 \%)$ and $83.2\% \pm 6.1 \% (\pm 17.9 \%)$ when analyzing low resolution mechanical data consisting of $64 \stimes 64$, $32 \stimes 32$, $16 \stimes 16$ and $8 \stimes 8$ vectors, respectively, see also Fig. \ref{fig:resolution}b).
Lower mechanical resolutions ($4\stimes4$) are too coarse for the two-dimensional autoencoder, and the reconstruction accuracy drops to $72.6 \% \pm 7.4 \% (\pm 17.9 \%)$ accordingly.
Confirming the results in section \ref{sec:results:temporal}, the reconstruction accuracy decreases to $89.9 \% \pm 4.3 \% (\pm 13.7 \%)$ (model 2Ds-E1s) and $87.3 \% \pm 5.0 \% (\pm 15.7 \%)$ (model 2Ds-F1s) with $32 \stimes 32$ and $16 \stimes 16$ vectors, respectively, if only static mechanical data is analyzed by the network, see also Fig. \ref{fig:resolution}c).
Remarkably, with three-dimensional mechanical data, the reconstruction accuracy remains the same at $2\stimes$ lower mechanical resolution, $95.7\% \pm 0.5\% (\pm 6.8 \%)$ (model 3Ds-B1) versus $95.7\% \pm 0.6\% (\pm 6.7 \%)$ (model 3Ds-A1). The accuracy decreases to $94.8 \% \pm 0.7 \% (\pm 8.1 \%)$ and $89.4\% \pm 1.4 \% (\pm 14.4 \%)$ at $4\stimes$ and $8\stimes$ lower mechanical resolutions (models 3Ds-B1 and 3Ds-F1), respectively, see Table \ref{table:networkarchitecture} and also Supplementary Movie 8.
The results demonstrate that autoencoders are very effective at interpolating from sparse data and that, accordingly, an autoencoder-based inverse mechano-electrical reconstruction approach is effective even when mechanics are analyzed at low spatial resolutions.

\subsection{Reconstruction of Active Stress (or Calcium Wave) Patterns from Mechanical Deformation} 

\begin{figure}
\centering
\includegraphics[width=0.46\textwidth]{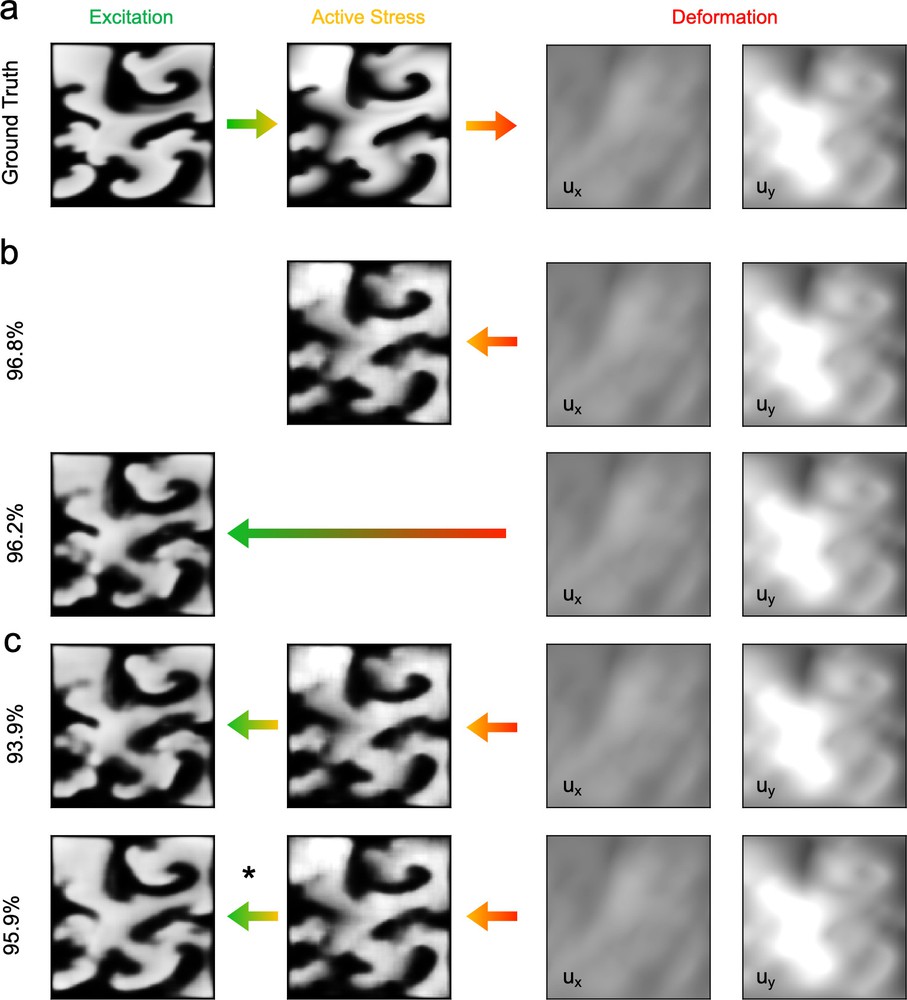}
\caption{
Reconstruction of active stress $T_a$ or excitation $V$ from mechanical deformation.
\textbf{a)} Forward excitation-contraction coupling converting an excitation wave pattern (left, n.u. [0,1]) into an active stress pattern (center, n.u. [0,0.5]), which leads to a corresponding elastic deformation of the tissue (right, displacement components $u_x$ and $u_y$ [-4,4] pixels).
\textbf{b)} Separate estimation of active stress (top) or excitation (bottom).
\textbf{c)} Indirect estimation of excitation from active stress pattern, which was estimated from deformation. Model trained with $T_a \rightarrow V$ (top) or $\widetilde{T}_a \rightarrow V$ (bottom, asterisk).
}
\label{fig:activestress}
\end{figure}

The neural network can be trained to estimate independently either electrical excitation $V$ or active stress $T_a$ from mechanical deformations, see Fig. \ref{fig:activestress}.
Estimating the active stress variable $T_a$, see eq. \ref{eq:modelTa}, from mechanics after the network (model 2Dt-A3) was trained with mechanical and active stress data, the reconstruction accuracy is $\langle \Delta T_a \rangle = 96.8 \% \pm 2.4 \% (\pm 3.2 \%)$, which is slightly (insignificantly) better than when $V$ is estimated with the same network, see Fig. \ref{fig:activestress}b) and table \ref{table:networkarchitecture}.
To allow the comparison, the active stress variable $T_a$ was scaled by a factor of $2$ before training (all $T_a<0.45$ [n.u.]) and normalized (with a fixed factor $\sim 0.45^{-1}$) before calculating the reconstruction accuracy.
Afterwards, the estimated active stress $\widetilde{T}_a$ can be used to indirectly estimate $V$, see Fig. \ref{fig:activestress}c), using a second network that performs the corresponding cross-estimation between the two scalar patterns $T_a \rightarrow V$ (accuracy: $97.9\% \pm 1.4\% (\pm 3.8 \%)$ with $n_d=1$ in input and output layer). If the second network was trained with ground truth values of $T_a$ and $V$ (top), the two subsequent estimations $\chi_t \rightarrow T_a \rightarrow V$ yield an overall reconstruction accuracy of $93.9\% \pm 4.6\% (\pm 12.2 \%)$.
If instead the second network was trained with estimations of $\widetilde{T}_a$ and ground truth values of $V$ (bottom, asterisk), the overall reconstruction accuracy after $\chi_t \rightarrow \widetilde{T}_a \rightarrow V$ slightly increases: $\langle \Delta V \rangle = 95.9\% \pm 3.7\% (\pm 9.6 \%)$.

\section{Discussion}
\label{sec:Discussion}
Our study demonstrates that autoencoders are a simple, yet powerful machine learning technique, which can be used to solve the inverse mechano-electrical problem in numerical models of cardiac muscle tissue.
We showed that neural networks with a generic convolutional autoencoder architecture are able to learn the complex relationship between electrical excitation, active stress and mechanical deformation in simulated two- and three-dimensional cardiac tissues with muscle fiber anisotropy, and can generate highly accurate reconstructions of electrical excitation or active stress patterns through the analysis of the deformation that occurs in response to the excitation or active stress, respectively. In the future, similar deep learning techniques could be used to estimate and visualize intramural action potential or calcium wave dynamics during the imaging of heart rhythm disorders or other heart disease, in both patients or during basic research. Given adequate training data, autoencoder-based convolutional neural networks and extensions or combinations thereof could be applied to analyze the contractile motion and deformation of the heart in imaging data (e.g. obtained with ultrasound \cite{Christoph2018} or magnetic resonance imaging) to compute reconstructions of electrophysiological wave phenomena in real-time. 
Our network is able to analyze a single volumetric mechanical frame and reconstruct a three-dimensional excitation wave pattern within that volume (containing $100 \times 100 \times 24 = 240,000 \approx 60^3$ voxels) in less than $20\,\text{ms}$, which suggests that performing the computations in real-time at a rate of $50$ volumes per second could in principle be achieved in the near future (e.g. with an Acuson sc2000 ultrasound system from Siemens using the matrix-array transducer 4Z1c, which provides volumes with $\sim 80^3$ voxels at that imaging rate).

We previously demonstrated - also in numerical simulations - that it is possible to reconstruct complex excitation wave patterns from mechanical deformation using a physics- or knowledge-based approach \cite{Lebert2019}. In contrast to the data-driven modeling approach reported in this study, the physics-based approach required a biophysical model to enable the reconstruction of the excitation from mechanics.
The biophysical model contained generic descriptions of and basic assumptions about the underlying physiological processes, including cardiac electrophysiology, elasticity and excitation-contraction coupling, and required the careful optimization of both electrical and mechanical model parameters for the reconstruction to succeed. 
In contrast, the data-driven approach in this study does not require knowledge about the biophysical processes or a physical model at all, which not only reduces computational costs, but also obviates constructing a model and selecting model parameters, which both could potentially contain bias or inaccurate assumptions about the physiological dynamics. Data-driven approaches generally circumvent these problems, but do require adequate training data that includes all or many of the problem's important features and lets the model generalize.
Our model was trained on a relatively homogeneous synthetic dataset and proved to be robust in its ability to reconstruct excitation waves from deformation in the presence of noise, at lower spatial resolutions, with arbitrary mechanical reference configurations, and in parameter regimes that it was not trained on. The robustness of the approach may indicate that similar deep learning techniques could be applied to solve the inverse mechano-electrical problem with experimental data, which is typically more heterogeneous, less systematic and of a lower quality overall. 
Nevertheless, in future research, one of the main challenges may be to generate training data that enables solving the real-world mechano-electrical problem with imaging data, and that, at the same time, captures the large variability and heterogeneity of diseases and disease states in patients.

In this study, we showed that autoencoder neural networks can reconstruct excitation wave patterns robustly and with very high accuracies from deformation, even though both the electrical excitation and the underlying muscle fiber orientation are completely unknown.
Nevertheless, at the same time, the autoencoder did not provide a better understanding of the relationship between excitation, active stress and deformation.
Similarly, we showed that the excitation wave pattern can be recovered, even though the underlying dynamical equations and system parameters that describe the relationship between electrics and mechanics are completely unknown to the reconstruction algorithm, but we did not gain insights into the mechanisms that underlie the reconstruction itself. 
Instead, the autoencoder learned the relationship automatically, in an unsupervised manner, and behaves like a 'black box' during prediction, a caveat that is typically associated with artificial intelligence.
It is accordingly difficult to assess whether it will be possible to develop a better understanding of the mutual coupling between voltage, calcium and mechanics in the heart using machine learning, and whether it will be feasible to predict under which circumstances machine learning algorithms may fail to provide accurate reconstructions. 
Hybrid approaches or combinations of machine learning and physics-based modeling \cite{Pathak2018Chaos} may be able to address these issues in the future.

We made a critical simplification in this study in that we assumed that the coupling between electrics and mechanics would be homogeneous and intact throughout the medium, and we did not consider electrical or elastic heterogeneity, such as scar tissue, heterogeneity in the electromechanical delay, the possibility of electromechanical dissociation, or fragmented, highly heterogeneous calcium wave dynamics during fibrillation. We aim to investigate the performance of our approach in the presence of heterogeneity and dissociation-related electromechanical phenomena more systematically in a future study.
For a more detailed discussion about potential future biophysical or physiology-related limitations, such as the degeneration of the excitation-contraction coupling mechanism or dissociation of voltage from calcium cycling during atrial or ventricular arrhythmias, we refer the reader to our previous publications \cite{Lebert2019,Christoph2018} and to related literature \cite{Wang2014,Omichi2003,Warren2007,Weiss2010,Voigt2012,Greiser2010,Greiser2014}.

\section{Conclusions}
\label{sec:Conclusions}
We provided a numerical proof-of-principle that cardiac electrical excitation wave patterns can be reconstructed from mechanical deformation using machine learning. 
Our mechano-electrical reconstruction approach can easily recover even complex two- and three-dimensional excitation wave patterns in simulated cardiac muscle tissue with reconstruction accuracies close to or better than $95\%$. 
At the same time, the approach is computationally efficient and easy to implement, as it employs a generic convolutional autoencoder neural network.
The results suggest that machine or deep learning techniques could be used in combination with high-speed imaging, such as ultrasound, to visualize electrophysiological wave phenomena in the heart.

\section{Supplementary Material}
{\bf Supplementary Figure:} Noisy mechanical deformation data analyzed by neural network autoencoder, see also Supplementary Movie 10. The inverse mechano-electrical reconstruction achieves high reconstruction accuracies even with noisy data (here shown for noise with a magnitude of 0.3), see also Fig. \ref{fig:noise}.
$\\$
$\\$
{\bf Supplementary Movies:} Supplementary Movies are available online or at \href{https://gitlab.com/janlebert/arxiv-supplementary-materials-2020}{gitlab.com/janlebert/arxiv-supplementary-materials-2020}.
$\\$
$\\$
\textbf{Supplementary Movie 1:} Reconstruction of two-dimensional chaotic electrical excitation wave dynamics from mechanical deformation in elastic excitable medium with muscle fiber anisotropy. First sequence: mechanical deformation that is analyzed by autoencoder. Second sequence: analyzed mechanical deformation (left) and reconstructed electrical excitation pattern (right). Third sequence: reconstructed electrical excitation pattern (left) and original ground truth excitation pattern (right).
$\\$
$\\$
\textbf{Supplementary Movie 2:} Reconstruction of three-dimensional electrical excitation wave dynamics from deformation in deforming bulk tissue with muscle fiber anisotropy. Left: Original electrical excitation scroll wave pattern. 
Center: Reconstructed electrical excitation scroll wave pattern using autoencoder model 3Ds-A1.
Right: Vortex filaments of original and reconstructed scroll waves.
$\\$
$\\$
\textbf{Supplementary Movie 3:} Electrical scroll wave chaos and  deformation induced by the scroll wave chaos shown separately for the top layer of the bulk, c.f. Fig. \ref{fig:results3D1}a).
$\\$
$\\$
\textbf{Supplementary Movie 4:} Three-dimensional reconstruction of dataset used in \cite{Lebert2019}. Left: Original electrical excitation scroll wave pattern. 
Center: Reconstructed electrical excitation scroll wave pattern using autoencoder model 3Ds-A1.
Right: Difference.
$\\$
$\\$
\textbf{Supplementary Movie 5:} Reconstruction with network model 2Dt-A3s using a temporal sequence of 3 mechanical frames to reconstruct 1 electrical frame showing estimate of electrical excitation wave pattern.
$\\$
$\\$
\textbf{Supplementary Movie 6:} Improvement of reconstruction with increasing training duration. Reconstructed excitation spiral wave pattern after 1, 2, 4, 6, 10 and 50 training epochs and comparison with ground truth.
$\\$
$\\$
\textbf{Supplementary Movie 7:} Reconstruction of two-dimensional focal excitation wave dynamics from mechanical deformation in elastic excitable medium with muscle fiber anisotropy.
$\\$
$\\$
\textbf{Supplementary Movie 8:} Reconstruction of three-dimensional electrical excitation wave dynamics from deformation at $4\times$ lower spatial resolution. Left: Original electrical excitation scroll wave pattern. 
Right: Reconstructed electrical excitation scroll wave pattern using autoencoder model 3Ds-D1.
$\\$
$\\$
\textbf{Supplementary Movie 9:} Reconstruction of two-dimensional electrical excitation wave dynamics from deformation at lower resolutions. 
Left: Mechanical deformation at $16 \stimes$, $8 \stimes$ and $4 \stimes$ lower spatial resolutions ($8 \stimes 8$, $16 \stimes 16$, $32 \stimes 32$ vectors) and original resolution ($128 \stimes 128$ vectors). Note that the mesh displays $9 \stimes 9$, $18 \stimes 18$, $36 \stimes 36$ lines.
Right: Reconstructed electrical excitation wave pattern and ground truth.
$\\$
$\\$
\textbf{Supplementary Movie 10:} Noisy mechanical displacement data (with magnitude 0.3). The noise is Gaussian distributed and was added independently onto the $u_x$- and $u_y$-components of the displacement vectors, see also Fig. \ref{fig:noise}.

\section{Funding}
This research was funded by the German Center for Cardiovascular Research (DZHK e.V.), Partnersite G\"ottingen (to JC).

\section{Acknowledgements}
We would like to thank S. Herzog and U. Parlitz for fruitful discussions on neural networks, and S. Luther and G. Hasenfu${\text \ss}$ for continuous support.

\section{Data Availability Statement}
The data that support the findings of this study are available from the corresponding author upon reasonable request.

\section{References}
\bibliography{references}

\end{document}